\documentclass[12pt,preprint]{aastex}
\usepackage{amsmath, amssymb, amsthm, amsfonts}
\usepackage{fullpage}
\usepackage{graphicx}
\usepackage{booktabs}
\usepackage{natbib} 
\def\lta{\,\raise 0.3 ex\hbox{$ < $}\kern -0.75 em
 \lower 0.7 ex\hbox{$\sim$}\,}
\def\gta{\,\raise 0.3 ex\hbox{$ > $}\kern -0.75 em
 \lower 0.7 ex\hbox{$\sim$}\,}

\usepackage{color}
\usepackage{xcolor}
\usepackage{hyperref}
\usepackage{breakurl}
\usepackage{amsmath}
\usepackage{graphicx}
\usepackage{verbatim}
\usepackage{booktabs}
\newcommand{\degree}{$^{\rm{o}}$} 
\newcommand{\be}{\begin{equation}}
\newcommand{\ee}{\end{equation}}

\begin{document} 

\title{Effects of Unseen Additional Planetary Perturbers on Compact Extrasolar Planetary Systems } 

\author{Juliette C. Becker$^1$ and Fred C. Adams$^{1,2}$} 

\affil{$^1$Astronomy Department, University of Michigan, Ann Arbor, MI 48109} 

\affil{$^2$Physics Department, University of Michigan, Ann Arbor, MI 48109} 

\affil{$\,$} 

\begin{abstract}

Motivated by the large number of compact extrasolar planetary systems discovered by the Kepler Mission, this paper considers perturbations due to possible additional outer planets. The discovered compact systems sometimes contain multiple transiting planets, so that their orbital angular momentum vectors are tightly aligned.  Since planetary orbits are susceptible to forced oscillations of their inclination angles, the highly aligned nature of these systems places constraints on possible additional (non-transiting) planets. If planets in the outer regions of these solar systems have sufficiently large mass or sufficiently small semi-major axis, they will induce the compact inner orbits to oscillate in and out of a transiting configuration. This paper considers the dynamics of the compact systems discovered to host five or more planets. In order to not perturb these systems out of a continually, mutually transiting state, additional planetary companions must generally have periastron $p>10$ AU. Specific constraints are found for each of the 18 planetary systems considered, which are obtained by marginalising over other orbital parameters using three different choices of priors for the companion properties (a uniform prior, a transit-inspired prior, and an non-transiting disk prior). A separate ensemble of numerical experiments shows that these compact systems generally cannot contain Jupiter-analogs without disrupting the observed orbits. We also consider how these constraints depend on system properties and find that the surface density of the planetary system is one of the most important variables. Finally, we provide specific results for two systems, WASP-47 and Kepler-20, for which this analysis provides interesting constraints. 
\end{abstract}


\section{Introduction}

In this work, we examine the effect of outer perturbing companions on compact systems of closely-packed planets (specifically, a subset of the multi-planet systems discovered by the Kepler spacecraft). This paper considers the effects of including hypothetical companions into the observed systems, with a focus on whether or not the entire system of planets remains in a mutually transiting configuration. The analysis thus considers the secular, dynamical, and transit stability of the systems, but does not provide any constraints on the long-term dynamical fate of the system or its formation history.  

Our knowledge of high planetary multiplicity systems originates largely from the Kepler mission \citep{systems2010, systems2011}. The Kepler mission has enabled population-level progress such as constraining the size distribution of exoplanets \citep{howard, fressin},  searching for and determining the abundance of rocky habitable exoplanets \citep{dressing, petigura, burke}, defining a transition zone between rocky and gaseous planets \citep{notrocky}, and much more additional work that was impossible before the era of large-scale transit surveys. One particularly interesting sub-population that has been found is a collection of high-multiplicity systems. More than forty planetary systems were found by the original Kepler mission to have four or more transiting planets, and more such systems continue to be found by the K2 mission \citep{fiveplanet}. 

Although the Kepler mission found many short-period planets, long-period planets are harder to find. Their transits are much less frequent and their radial velocity signals are smaller than for planets closer to the star. Finding long-period companions to existing Kepler systems with any number of known planets can be approached from two directions: observational searches and theoretical constraints. 
Observational searches that have met with success at finding potential long-period companions include re-analysing legacy data \citep{legacy, longperiod,uehara}, conducting follow-up radial velocity searches \citep{friends_rv} or searching using adaptive optics \citep{ao13,ao16,ao162}.

Theoretical searches, in contrast, can inform on what populations of unseen planets could exist in principle. These studies can be either analytic or numerical. The analytic approach is by necessity limited in scope, as analytically tractable equations cannot encapsulate the full behaviour of a complex planetary system. As an example, \cite{lai} generalised the secular approximation for the behaviour of systems with more than two planets, and found that perturbing companions can indeed excite the inclinations of the orbits in the inner systems. 

In general, the secular approximation is an efficient, time-saving technique (see also \citealt{secular1}, \citealt{secular2}, \citealt{secular3}, and many additional recent papers) which is often applied to this problem. However, to evaluate the effects of massive, outer perturbers on an inner compact system, one must use full N-body numerical simulations. Multi-planet systems are often highly chaotic, so that many realisations of the systems must be considered to fully evaluate their dynamics. The numerical approach can thus test specific systems and determine probabilities of varying potential outcomes. \citet{chelsea} used N-body experiments to test the stability of super-Earth systems in the presence of a companion exterior to 1 AU, and determined that a majority of super-Earth systems are destabilised by the presence of such a perturbing companion. \citet{priors_mustill} performed a complementary set of numerical experiments, but found the same destabilising effect (albeit at a lower rate). 

In this work, our goal is to test not only the stability of observed Kepler multi-planet systems, but also to map their transiting behaviour. \citet{petrovich} found that for systems where tightly packed inner planets have significant eccentricities or inclinations, the excitation of those orbital elements must occur before the planets attain their tightly packed configurations. This finding suggests that a tightly packed system with all of its planets observed to be transiting would not generally be expected to attain a non-transiting configuration over secular timescales, although this stability may not hold over the age of the system. Another previous study \citep{volk} found that currently observed multi-planet systems may be the remnants of tightly packed compact systems, which may have lost planets over time through dynamical instabilities and collisions over the history of the system. On the other hand, \citet{johnm} found that these systems may be dynamically stable over spans of time much longer than secular timescales. 

A comprehensive numerical analysis requires an average of 5000 -- 8000 CPU hours on standard processors for each planetary system under investigation. For this reason, and others, the behaviour of the Kepler multi-planet systems in the presence of extra companions has not yet been evaluated numerically. For these compact planetary systems, this investigation thus provides a picture of the transiting behaviour of the inner planets in the presence of an extra companion. 

%For example, \citep{carrera} determined that if companions have large eccentricities, they can destabilise the dynamical stability of an interior system of planets. 
%\citep{moveplane} found that an exterior companion can excite misalignments between a system of planets and its host star.

As such, this work builds upon an earlier contribution \citep{paper1}, where we examined the possible self-excitation of inclination in a collection of the multi-planet systems discovered by the Kepler mission. These Kepler multi-planet systems are generally tightly packed, with four or more planets orbiting within $\sim0.5$ AU. Self-excitation of inclination occurs when planets that are a part of such a tightly packed system trade angular momentum among their orbits. The end result is that one or more planets could have inflated inclinations at any given time. Over secular timescales, the particular planets that are excited to higher inclinations may change, and the width of the effective ``plane'' containing the planets may also vary. As a result of such interactions, a system where all planets start in a mutually transiting configuration from our line of sight could evolve such that one or more planets leave the transiting plane at later times \citep[a related treatment of this problem can be found in][]{corbits}. 

For completeness, we note that the system could also evolve to a configuration where all of the planets are observed in transit from a different line of sight. 
The movement of planets in and out of the transiting plane could also excite large observed obliquities, as planets move out of the plane aligned with the stellar spin axis of the host star. Multiple authors \citep{gongjie_tides, mazeh, tim} have found that the systems with multiple planets have lower obliquities. 

Using a combination of secular and numerical analyses of the multi-planet systems observed by Kepler, previous work determined that self-excitation is not generally extreme enough to cause most of the observed systems to attain non-transiting configurations. For the sake of definiteness, we call the state of being in a continually mutually transiting configuration ``CMT-stability". Note that systems that are ``CMT-unstable'' are usually dynamically stable, in that they retain all of their planets (see also \citealt{corbits} for a more detailed discussion). However, for a more general set of systems --- those motivated by the observed sample, but with a wider range of allowed properties ---  self-excitation can have a greater effect, leading to potentially CMT-unstable systems. In other words, planetary systems that are nearly the same as those observed, but with slightly different specific orbital elements, can oscillate in and out of transit. This effect --- changes in the observability of a system over time due to dynamical interactions --- has received much recent attention \citep[see][]{ballard14}, including some studies of the effect of extra, unseen, perturbing bodies \citep[prepared simultaneously with this work; see][]{lai, priors_mustill, priors_hansen}. This present paper carries previous work forward by performing a more robust ensemble of numerical simulations for the systems with the highest multiplicity (with the caveat that numerical limitations prevent us from analysing the entire Kepler multi-planet sample in this way). This study also examines the effects of different priors on the end results by choosing three different versions of the priors. Finally, we provide predictions for specific systems (such as Kepler-20 and WASP-47). 

In our previous paper, we considered the systems to contain only the bodies observed thus far. It is unlikely that our observations are complete, and so it is useful to examine the effects of a perturbing body (giant planet or brown dwarf) on each compact, multi-planet system, using the same basic methodology that we did for the compact systems without perturbing bodies. Moreover, sufficiently distant companions are likely to be found outside the orbital plane of the inner system: the original molecular cloud cores that produce star/disk systems often have a range of angular momentum vectors and this complication, along with dynamical evolution, can often lead to companion orbits that are inclined (see \citealt{barclay}, \citealt{chris1}, and references therein). 

This study derives statistical limits that constrain the presence of companions in the observed multi-planet Kepler systems, given that we see them in transit today. 
In this work, we place these limits by performing a large number of computationally-intensive simulations for 18 of the observed Kepler multi-planet systems. In Section \ref{LOS}, we discuss our numerical techniques and some typical results that characterise the effects of perturbing bodies for individual systems. Note that it is not feasible to carry out a detailed numerical analysis for every compact system that will be discovered. Section \ref{sec:limits} presents limits on potential unseen companions for the sub-sample of multi-planet systems considered in this work. These results provide a general picture of the CMT-stability for observed systems with planets in compact orbits and suggest methods for predicting the companion status of such systems. Section \ref{sec:insights} presents specific results for the dynamically interesting systems Kepler-20 and WASP-47. The paper concludes in Section \ref{sec:conclude} with a summary of our results and a discussion of some limitations of this analysis.

\section{Evaluating the Effect of Unseen Companions \\ on the Observed Kepler Multi-Planet Systems}
\label{LOS}
%Statistical limits on companion presence in Kepler multi-planet systems, given low stellar spin - planetary orbit obliquities 
%explain how we did it
%Include a table with the radius outside of which it is unlikely to have a companion. Maybe this should be a plot or two instead...

The compact, multi-planet systems discovered by Kepler are remarkably stable in their currently observed transiting configurations, as long as there are no extra companions in the systems. If an additional body (giant planet or star) is introduced, however, the behaviour of the currently observed planets could be significantly altered over secular (and longer) timescales. Sufficiently large and/or close perturbing bodies could lead to the inner system becoming either dynamically unstable or CMT-unstable. Both of these scenarios would lead to a complete system with different properties than these observed by Kepler. Note that the perturbing bodies themselves could move in and out of transit with time, and would transit with low probability due to their large orbital separations. Notice also that the observed systems tend to have regularly-spaced orbits \citep{spacing} with no large gaps where non-transiting planets could reside. We thus expect any additional planets to generally lie outside the observed compact systems. 

Since the multi-planet systems were indeed discovered by Kepler, we can rule out the presence of companions in these systems that would disrupt their CMT-stability on short timescales. In addition, companions that disrupt CMT-stability on longer timescales are unlikely. For example, in all of the Kepler multi-planet systems considered in this work, a brown dwarf orbiting at 0.5 AU would disrupt the orbits of the inner planets, so that the inner system would fail to transit continually. The systems would thus be ``CMT-unstable'' and planetary systems of this type would not have been discovered by Kepler in their observed configurations. However, for a diametrically different companion type --- say, a super-Earth at $a$ = 700 AU, analogous to the proposed Planet 9 in our own solar system \citep{planet9} --- none of the systems in our sample would be disrupted from their current transiting configurations. It is clear that for every compact multi-planet system, there is some regime of ``acceptable" companions, which could very well exist in the observed systems as they do not alter the orbits of the compact system planets, and some other regime of ``unacceptable" companions, which lead to the inner system being CMT-unstable. The goal of work is to determine these limits using numerical N-body integrations of the observed Kepler multi-planet systems. 

\subsection{The Necessity of N-body Integrations}
\label{neednbody}
In previous work, we used Laplace-Lagrange secular theory to evaluate the CMT-stability of the observed Kepler systems with four or more planets (\citealt{paper1}; see also \citealt{lai}). In that context, secular theory is an appropriate approximation both because the planets attain fairly low eccentricities and inclinations through the natural orbital evolution, and because the systems are (as observed) dynamically stable. These systems do not experience orbit-crossing, scattering, or other dynamical complications that depend on the mean motions and could cause a system to change its orbital configuration. 

\begin{figure}[htbp] %  figure placement: here, top, bottom, or page
\centering
\includegraphics[width=6in]{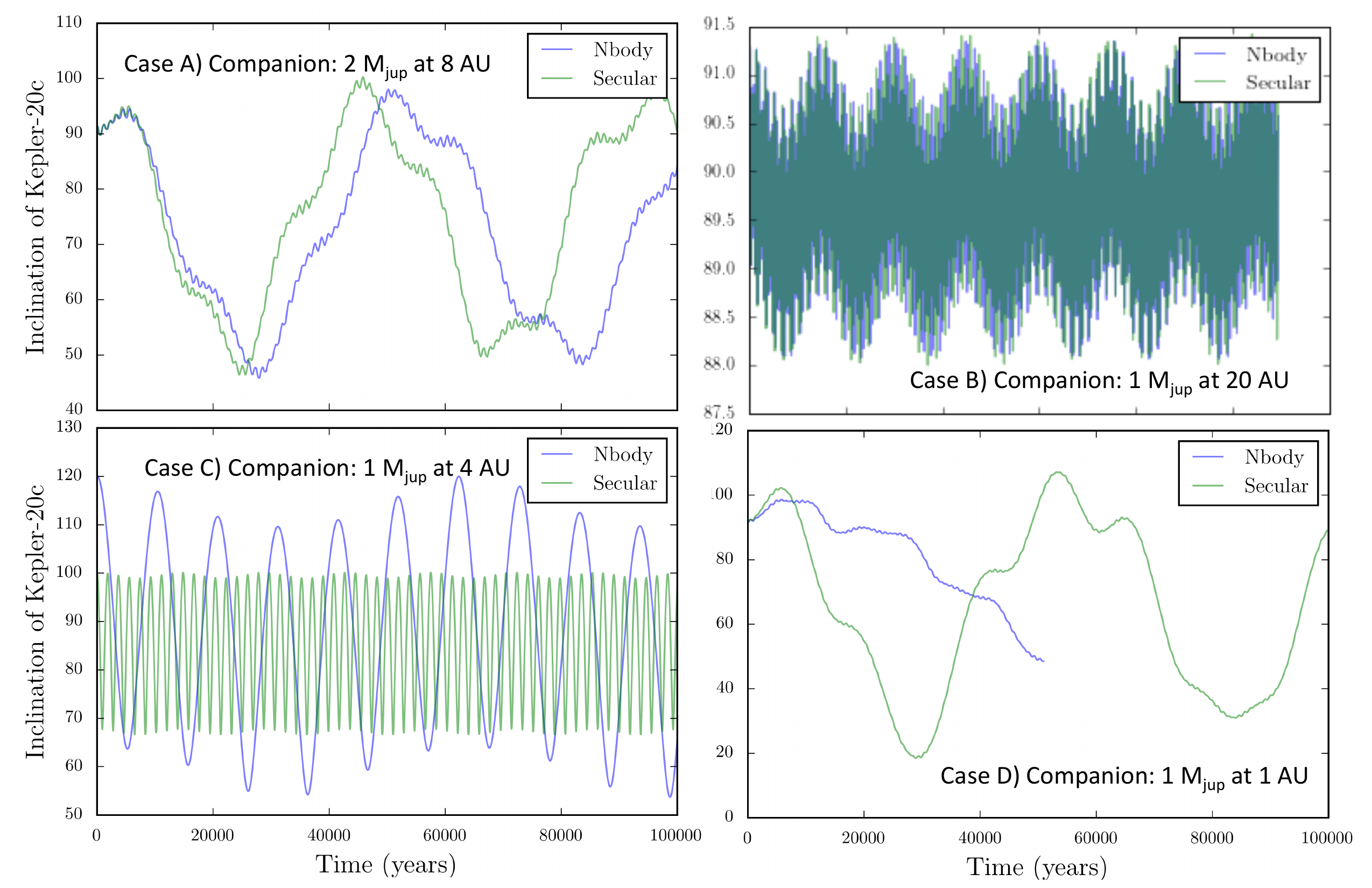} 
\caption{Four different companion types and their effects on the evolution of Kepler-20c, as computed using both secular and numerical methods. Case A shows the well-known effect of the potential inconsistency in the period of oscillations, which is important but unlikely to affect our results. Case B shows a case with good agreement between the two methods, and demonstrates that for large orbital separation companions, secular theory may be a good approximation. Case C shows differences in both the period and amplitude of the oscillations, an effect large enough to potentially skew results. Case D shows an example where the numerical integration identifies a dynamical instability that would be ignored in the secular approximation. For all cases, the realisations plotted were chosen from more than 4000 integrations, selected for illustrative purposes. }
   \label{new_stab_fig}
\end{figure}

When adding a perturbing companion to such a system, the opportunity arises for all of these mean-motion-dependent events to have significant effects on the evolution of the system. To illustrate this behaviour, Figure \ref{new_stab_fig} shows a selection of such effects for one particular planet (Kepler-20c), and illustrates the (at times) significant discrepancy between numerical N-body evolution and secular evolution.  The integrations shown in Figure \ref{new_stab_fig} are drawn from the sample constructed for this work (which will be described in depth in Section \ref{nbodysims}). The secular analogues were generated using the same starting parameters that were used in the N-body integrations. The plots cover only $10^{5}$ years for ease of viewing the relevant oscillations. 

One major, well-known, unavoidable difference between secular theory and N-body integrations is the timescales of periodic evolution (this effect is explained, using Jupiter and Saturn as examples, in chapter 7 of \citealt{md}).  This effect is illustrated in Case A of Figure \ref{new_stab_fig}, which shows the evolution of Kepler-20c in the presence of a 2 $M_{\rm jup}$ companion at 8 AU. In this case, the evolution proceeds similarly in both secular and N-body cases, but the orbital elements evolve on somewhat different timescales. This kind of deviation will not compromise our results to a large degree, as it does not change the amplitude of the oscillations. Although there may be slight deviations when simultaneous transits occur (between multiple inner planets, each evolving on different timescales), this effect will likely be small. Moreover, these effects are unlikely to influence CMT-stability. 

Case B in Figure \ref{new_stab_fig} shows the ideal  situation in which secular theory can be used to approximate the dynamics with high fidelity. In this case, a 1 $M_{\rm jup}$ companion in introduced at 20 AU. The orbits of the perturber and of the inner system are sufficiently spatially separated that no additional effects arise due to mean motions. In this case, all of the relevant dynamical variations are encapsulated by the secular approximation. 

In contrast, Cases C and D in Figure \ref{new_stab_fig} demonstrate the limitations of secular theory. Case C shows the effect of a 1 $M_{\rm jup}$ companion at 4 AU. Not only are the periods of the inclination oscillations different, as seen in Case A, but the amplitude of the oscillation shows a significant difference between the two calculations. This difference can be caused by a variety of factors which are not included in the secular approximation (including resonance-driven boosting of eccentricity or inclination, etc.). Since CMT-stability depends on the amplitude of inclination oscillations, a deviation of this magnitude will lead to different conclusions derived from using each method. 

Case D shows the effect of a 1 $M_{\rm jup}$ companion at 1 AU. Note that the outer orbital radius of the planets in the observed Kepler-20 system is 0.35 AU, so this companion orbit is a factor of three larger than that of the observed system. This case illustrates another major imperfection in secular theory. Here, the inner system of planets becomes dynamically unstable due to the companion: In this particular integration, the eccentricities are increased to such an extent that orbit crossing occurs (but instability can occur in many other ways). This instability is plainly evident in the numerical integrations, but the secular theory is insensitive to such effects (which depend on where a planet is on its orbit when the orbits cross). The numerical integrations are thus necessary to evaluate whether the planets survive or not. 

One additional limitation of secular theory that is not obvious from Figure \ref{new_stab_fig} is its treatment of semi-major axis. In secular theory, the semi-major axes of the interacting planets are fixed and only the evolution of other orbital properties (such as inclination, as plotted above, or eccentricity) is allowed to vary. On the other hand, numerical N-body simulations allow the semi-major axes to evolve. If the semi-major axes are expected to evolve (which would occur in scattering interactions with the perturber, for example), then it is necessary to use numerical techniques. 

Although the secular approximation is a tremendously useful tool for making dynamically difficult problems tractable without significant investments of CPU time, it is insufficient for the particular problem considered in this paper. Moreover, the secular theory fails for the regime of parameter space for which the planetary systems are dynamically active, i.e., the regime of interest in this study. As a result, we must turn to numerical methods.  

\subsection{Numerical methods}
\label{nbodysims}
%introduce idea
To evaluate the effects of unseen companions on the Kepler multi-planet systems, we use numerical N-body simulations to evaluate the dynamics and stability of the observed systems on secular timescales. These numerical integrations are carried out using the N-body code \texttt{Mercury6} \citep{m6}.\footnote{Note that a numerical treatment is required for reasons explained in Section \ref{neednbody}.} Because the systems are chaotic, and because we need to consider a range of possible companions, many integrations of each system must be carried out. Toward this end, we use a Monte Carlo technique to generate multiple realisations of each compact multi-planet system discovered by Kepler, and introduce a perturbing companion with randomly chosen orbital elements. We then evaluate the CMT and dynamical stability of the system for each such realisation. The unseen companions are assumed to have orbits exterior to the observed compact Kepler systems. The distributions of the companion masses and other properties are described below. 

% explain sim parameters

% explain relevant priors
\subsubsection{Priors} % might not need own subsection
\label{masses}
Not every orbital parameter describing the planetary systems we consider can be measured. In order to complete the specification of the initial conditions for the N-body integrations, we have to determine the remaining parameters, both the the observed planets and for the unseen companion. These distributions of parameters -- priors -- are described below. Note that the priors for the observed planets are a means to specify unobserved properties of the system, e.g., setting the planetary mass when the radius is measured. The priors for the putative companions are less constrained, as the additional bodies could have a wide range of orbits and other properties. 

The orbital elements for the existing planets were drawn from observational priors \citep[for the planets in our sample, these priors are found in][we use the values available as of January 2016]{systems2010, systems2010b, systems2011, systems2012, systems2013, systems2013b, systems2014, systems2014b, systems2014c}. When observational priors do not exist (for example, for most planets, inclination has not yet been observationally determined), we draw from a prior chosen to be representative of the possible values of that parameter. We use the same methodology as in \citet{paper1} to choose these values, and a more lengthy description of the choices made can be found in that paper, but the most important choices will be described with brevity here. 

\textbf{Planetary Masses.} Photometric light curves yield an excellent measure of planetary radii, but they do not provide direct measurement of the planetary masses. For this reason, it is necessary to use a mass-radius relation to choose starting masses for the simulations. The Wolfgang relationship from \citet{angie} provides a probabilistic mass-radius conversion function for planets in the range $R_P=1.5-4 R_{\rm{\oplus}}$. For planets smaller than this lower limit, we use the relation from \cite{weiss}. For planets larger than the upper limit, we use a characteristic gas giant density generated by a Gaussian draw from the observed gas giant densities \citep[as done in][]{fiveplanet}. 

\textbf{Orbital Inclinations.}  Mutual inclinations are generally parametrised by a Rayleigh distribution with some width \citep{2009ApJ...696.1230F, lissauer, fangmar, ballard14}, which is typically taken to lie in the range 1\degree -- 3\degree. For this application, we use a simple Rayleigh distribution with width 1.5\degree\ \citep{hillstab1} for planets without measured inclinations, with the constraint that all planets in the inner system must initially be transiting. 

\textbf{Additional Companion Properties.} Because we do not fully understand the population of long-period planets, the priors for our injected companions can be chosen in a variety of ways. This regime of parameter space, with semi-major axes $a>1$ AU, is not fully sampled observationally, and different techniques (RV, transit, direct imaging, micro-lensing) each have their own biases and limitations. 

In previous work concerning the stability of the Kepler systems, a variety of priors were used. For example, \citet{priors_mustill} used priors representing stellar and planetary companions, and found that not only are the dynamical instability rates are different between the two cases, but so too is the amount of inclination excited. \citet{priors_hansen} also performed numerical experiments that involved adding a perturbing body while using delta function priors and multiple trials, which provides a description of the average behaviour of the system for each companion type, but does not explain how susceptible results are to small changes in companion type. 

In this paper, we use three sets of priors for our population of perturbing bodies. The goal is to not only describe the behaviour of the inner systems in the presence of these extra bodies but also to determine the differences in computed stability thresholds using the different priors. As a result, for each observed system, we construct three samples of 2000 injected perturbing bodies by re-sampling our 4000 integrations with the following priors:

\begin{itemize}
\item \textbf{Transit-inspired prior.} The transit-inspired prior is intended to test the behaviour of perturbing bodies that may have formed in a plane with the inner system of planets, and remained roughly coplanar (see Section \ref{WASP47} for an example of a system that may have done this). For these companions, we choose a mass from a log-uniform distribution between 0.1 and 10 $M_{\rm jup}$. For inclination we use a Rayleigh distribution with a width of 3 degrees (which corresponds to the largest width given in recent papers that parametrise inclinations in this way; see \citealt{fangmar}, \citealt{ballard14}, and \citealt{hillstab1}). The orbit of the perturbing planet is thus close to the plane(s) of the inner system. For this prior, we choose eccentricity from a beta distribution with shape parameters $\alpha = 0.867$ and $\beta = 3.03$ \citep[an observationally motivated distribution derived in][]{kipping1}, and choose the argument of periastron according to the asymmetric, sinusoidal distribution given in that same paper.
\item \textbf{Uniform prior.} The uniform prior is intended to explore the entire parameter space that could potentially be populated by unseen companions.  The semi-major axes are uniformly distributed in the range $a=1-30$ AU. The planetary masses are uniformly distributed in the range $M_p=0-10$ $M_{\rm jup}$. The inclination drawn from a uniform distribution with range $i=0-90$\degree. The eccentricity is drawn from a beta distribution, whereas the argument of pericentre has a corresponding asymmetric distribution with a sinusoidal prior (see \citealt{kipping1}). Finally, the longitude of the ascending node $\Omega$ is taken to be uniform over the range $0-360$\degree.
\item \textbf{Non-transiting disk prior.} The non-transiting disk prior is intended to mimic the population of planets (and brown dwarfs) that could exist in these systems but be undetectable via transit methods. A large number of such planets have been discovered in existing systems (see, for example, \citealt{systems2014c}; note that the non-transiting planets in these systems are generally exterior to the transiting system). As in the other two populations, we draw the eccentricity and argument of periastron from the distributions given in \citet{kipping1}. We draw the companion mass from a distribution uniform in log space, ranging between 0.1 and 10 $M_{\rm jup}$. We draw the inclination from a uniform distribution between 60 and 90 degrees, representing a 0 -- 30 degree misalignment between the extra body and the plane of transiting planets. This 30 degree width is based on the maximum misalignment expected due to  variations in the angular momentum direction between molecular cloud cores and their forming circumstellar disks \citep{goodman,cores}. As discussed in \citet{barclay}, there is no expectation of correlation within this range, and thus we use a uniform distribution in inclination (allowing up to 30 degrees of misalignment). Finally, we choose the semi-major axis from a uniform range between 1 and 30 AU. 
\end{itemize}

\begin{figure}[htbp] %  figure placement: here, top, bottom, or page
\centering
\includegraphics[width=3.4in]{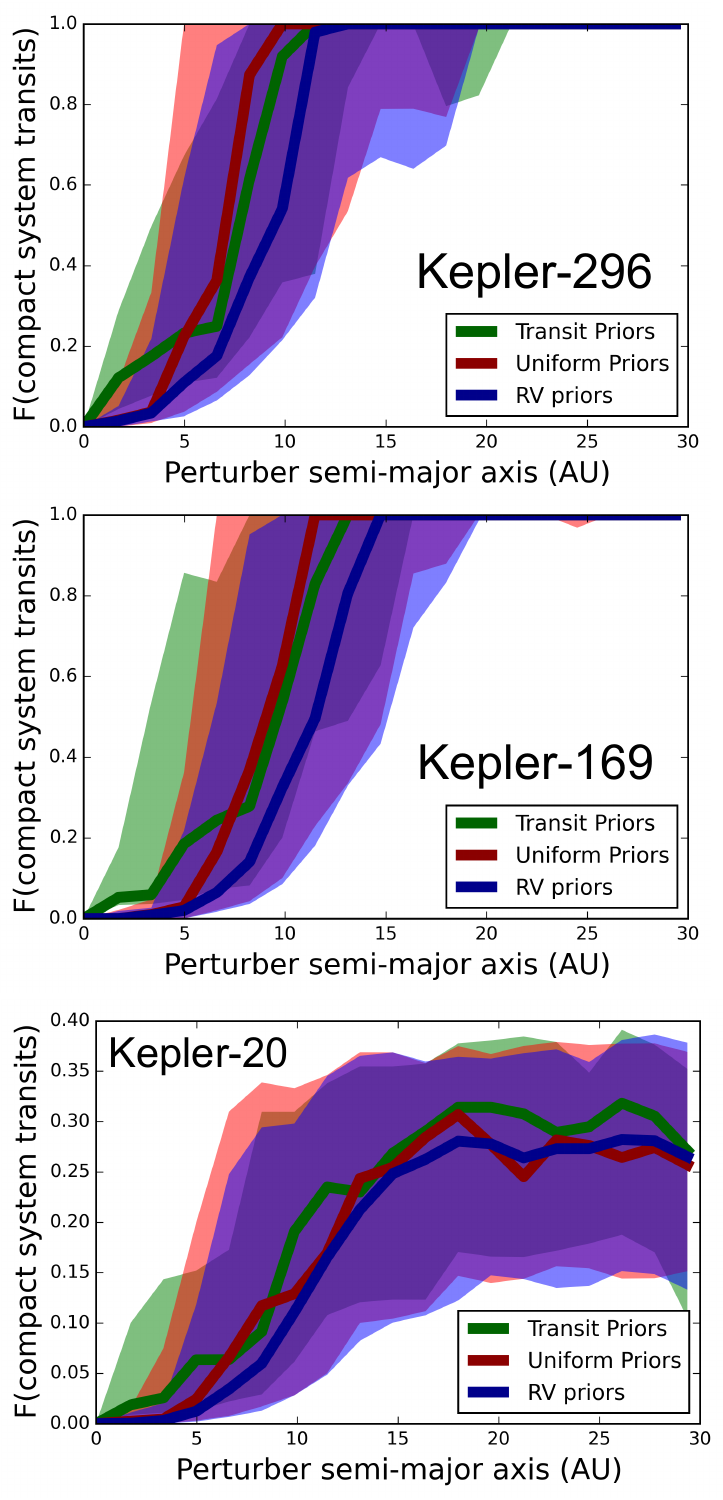} 
\caption{The three different priors for the unseen companion lead to slightly different results for CMT-stability, as shown here for three representative systems. As expected, the transit prior (which has a much narrower range of allowed inclinations than the other two choices) tends to have a larger fraction of systems that are CMT-stable at close distances to the star. However, the difference is not as large as might be expected, due to the width of the Rayleigh distribution used as a prior for inclination.  }
   \label{three_priors}
\end{figure}

In all three cases, the uniform sampling over semi-major axis does not bias our conclusions because we marginalise over semi-major axis (and later, periastron distance) in our results. We used the uniform prior to choose the orbital elements of all planets in the 4000+ trials (per system). To construct populations for the other two prior types, we resampled those initial 4000+ integrations and supplemented them with additional integrations for the Transit-inspired prior so as to have 2000 integrations for each prior. The results of a comparison between the three prior types can be seen in Figure \ref{three_priors} for a selection of example systems.

Figure \ref{three_priors} shows that the three choices of priors for the companion lead to similar results. The three systems plotted in this figure represent two typical systems (Kepler-296, Kepler-169) and one system with particularly large CMT-instabilities (Kepler-20). For all three considered systems, the fraction of realisations that remain CMT-stable are a well-defined and increasing function of separation, measured here through the semi-major axis of the companion. The ``transit prior'', which has the lowest set of inclination angles for the companion, leads to a larger fraction of CMT-stable systems at $a\sim5$ AU, but produces remarkably similar results at larger separations. In any case, for all three priors used here, the overall trend and extent of the populated region is roughly the same, even though the subtleties of the slopes may change between prior types. The differences between the systems are greater than the differences between the prior choices for a single system, so we are confident that the behaviour exhibited in our simulation results is attributable to the exoplanetary systems themselves, and not to our choice in prior. 

\begin{figure}[htbp] %  figure placement: here, top, bottom, or page
\centering
\includegraphics[width=3.4in]{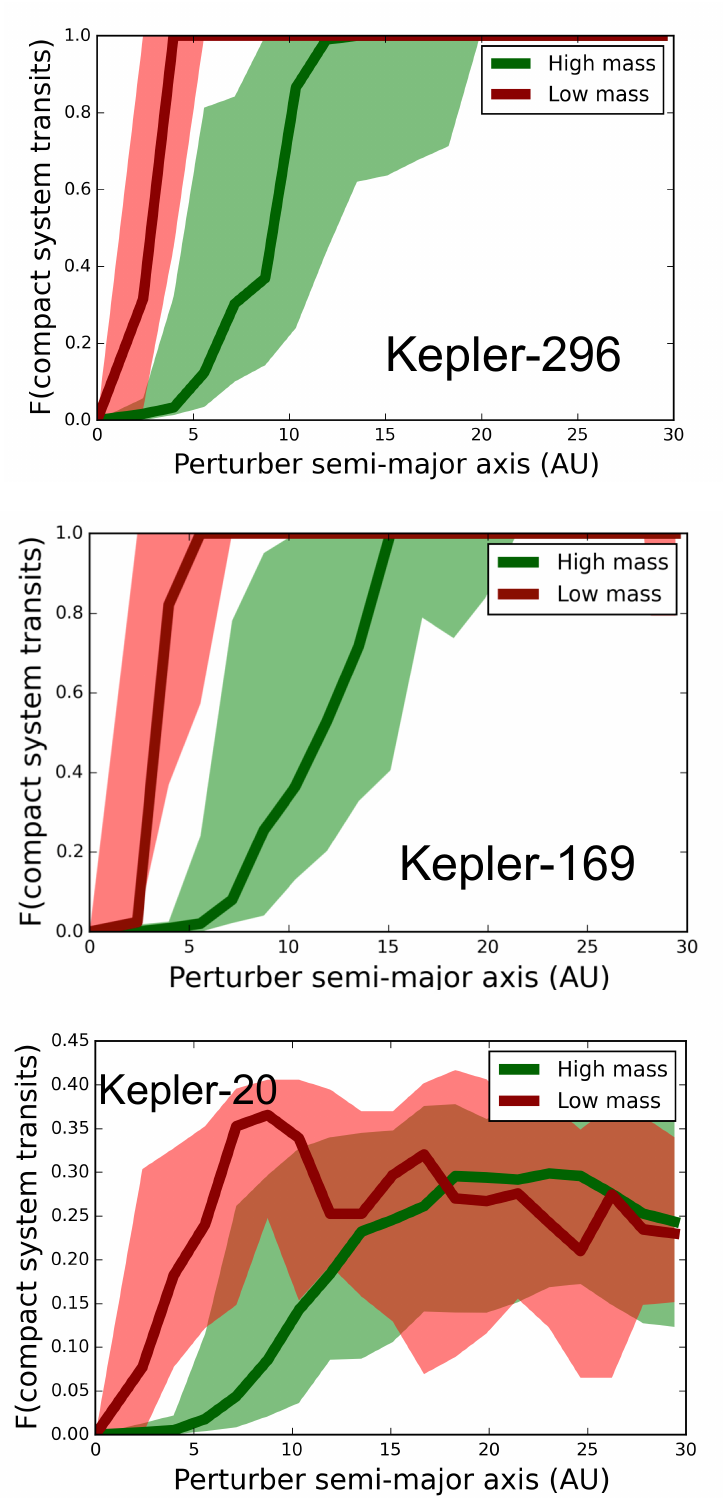} 
\caption{For the same three systems considered in Figure \ref{three_priors}, this figure shows the CMT-stability curves for systems with high-mass (green) and low-mass (red)  perturbers. The high-mass companions cause systematically more CMT-instability of the inner system. Analogous plots for eccentricity and inclination (not shown here) do not show an obvious difference between the high- and low-quantity populations.  }
   \label{two_mass}
\end{figure}

Figure \ref{two_mass} shows the CMT stability curves for the same three systems considered in Figure \ref{three_priors}. Instead of evaluating the difference of each prior type, the two curves here show the difference between the high- and low-mass (from the uniform prior, where high-mass is taken to be greater than 2 M$_{\rm{jup}}$ and low-mass is taken to be masses less than or equal to that value) values for the companion's mass. The low-mass companions are lead to a lower amount of CMT-instability in the inner system. 

In Figure \ref{three_priors} and Figure \ref{two_mass}, we plot the CMT-stability fraction against the semi-major axis of the perturber. This choice was motivated by our choices of priors: we naively sampled uniformly in semi-major axis with the intention to explore perturbers at all orbital radii, with the results marginalised over our three prior choices. Analogous plots to Figure \ref{two_mass} for eccentricity and inclination demonstrate significantly less variation between the high and low value populations for those quantities. As a result, we choose to plot the CMT-stability fraction against perturber periastron distance for the remainder of this work. Periastron is a more physically illustrative value than semi-major axis, since it describes the minimum distance attained between the perturber and the planets of the inner system.

% Requires the booktabs if the memoir class is not being used
\begin{table}[htbp]
   \centering
   %\topcaption{Table captions are better up top} % requires the topcapt package
   \begin{tabular}{@{} ccccc @{}} % Column formatting, @{} suppresses leading/trailing space
      \toprule
      \multicolumn{3}{c}{\textbf{Priors for compact inner system of planets}} \\
      \cmidrule(r){1-3} % Partial rule. (r) trims the line a little bit on the right; (l) & (lr) also possible
      Orbital Element    & Prior  & Reference \\
      \cmidrule(r){1-3} % Partial rule. (r) trims the line a little bit on the right; (l) & (lr) also possible
      semi-major axis, $a$   & observational limits  &  (various$^{2}$)\\
      planetary radius, $R_{p}$   & observational limits  & (various)  \\
      planetary mass, $M_{p}$   & step function, converted   & \cite{weiss}   \\
     &  from measured radius & \cite{angie}   \\
      inclination, $i$   & mutual inc. from Rayleigh  & \cite{hillstab1} \\
      &  distribution with width 1.5$^{\rm{o}}$   &   \\
      argument of pericentre, $\omega$   & uniform on $(0^{\rm{o}}, 360^{\rm{o}})$  &   \\
      longitude of ascending node, $\Omega$   & uniform on $(0^{\rm{o}}, 360^{\rm{o}})$  &   \\
       eccentricity $e$   & uniform on (0, 0.1) &  \\

    %  \\
     
      %\multicolumn{3}{c}{\textbf{Priors for outer perturbing companion}} \\
 %     \cmidrule(r){1-3} % Partial rule. (r) trims the line a little bit on the right; (l) & (lr) also possible
   %   Orbital Element    & Transit Prior & RV Prior & Uniform Prior  & Reference \\
 %     \cmidrule(r){1-3} % Partial rule. (r) trims the line a little bit on the right; (l) & (lr) also possible
 %     semi-major axis, $a$   & uniform on (1 AU, 30 AU)  &   \\
 %     planetary mass, $M_{p}$   & uniform on $(0 M_{\rm{jup}}, 10 M_{\rm{jup}})$  &   \\
  %    inclination, $i$   & uniform on $(0^{\rm{o}}, 90^{\rm{o}}) $  &   \\
  %    longitude of ascending node, $\Omega$   & uniform on $(0^{\rm{o}}, 360^{\rm{o}}) $ &   \\
  %     eccentricity $e$   & beta distribution with $\alpha=0.867$, $\beta=3.03$ & \citet{kipping1} \\
  %    argument of pericentre, $\omega$   & asymmetric, sinusoidal prior  & \citet{kipping1}   \\
      \bottomrule
   \end{tabular}
\caption{Priors for the Monte Carlo realisations of each planet in the observed Kepler multi-planet systems. Each system is comprised of the inner planets discovered by Kepler and a single outer, perturbing companion (whose orbital parameters are chosen using the three priors described in Section \ref{masses}). Although mass measurements do not exist for most of these planets, the radii of the observed planets are derived from the transit light curves and stellar radii found in the literature. From these radii, we use the conversion procedures summarised in \citet{paper1}, which use relations from \cite{angie} and \cite{weiss} to estimate planetary mass for each realisation, which results in orbital parameters for each of the compact system planets in each studied system.  (2) We downloaded best-fit orbital parameters from exoplanets.org as of January 2016, and updated when needed with parameters from \citealt{systems2010, systems2010b, systems2011, systems2012, systems2013, systems2013b, systems2014, systems2014b, systems2014c}. }
   \label{tabone}
\end{table}

\subsubsection{Simulation Parameters}

Every realisation of each system requires drawing the orbital elements for each planet, a process which has been described in Section \ref{masses}. Once starting parameters for each planet have been chosen, we check for the Hill stability of the inner system of planets (ignoring the outer perturber) before beginning the computation for the realisation. If the initial conditions for the inner system of compact planets are not Hill stable, we discard that realisation (as it does not give us useful information about how an outer companion affects the behaviour of the inner system). We check for Hill stability, a criterion for stability that has the form 
\begin{equation}
\frac{a_{out} - a_{in}}{R_{H}} > \Delta_{crit}\,,
\end{equation}
where $\Delta_{crit} = 2 \sqrt3$ is the critical separation for adjacent planets, and $R_{H}$ is the mutual Hill radius, which is defined as 
\begin{equation}
R_{H} \equiv \left( \frac{M_{in} + M_{out}}{3 M_{c}} \right)^{1/3} \frac{a_{in} + a_{out}}{2} \,.
\end{equation}
Since we are considering systems with more than three planets, we also require that $\Delta_{inner} + \Delta_{outer} > 18$ for each pairing of inner planets \citep[as in][]{hillstab1, ballard14}. These initial conditions are intended to screen out systems where the inner system is dynamically unstable on its own, even without the perturber, due to unlucky draws from their observational priors. Since such unstable systems would not reflect the action of the perturber, these realisations are removed before the numerical integrations are run. Such cases are rare: Among the tens of thousands of realisations run, only a handful were discarded because of their failure to meet these constraints. 

After the starting parameters for all bodies in the system are chosen and the inner system is confirmed to be Hill stable, we integrate the realisation forward for $10^{7}$ years. If, during this time, any planets are ejected from the system, collide with the central body, or undergo a close encounter within 3 mutual Hill radii of another planet, we stop the integration and consider the system to be disrupted, and thus dynamically unstable for our purposes \citep[this approach is consistent with the criteria used in other work such as ][]{hillstab1}. We do not consider spin or tidal effects, as additional evolution due to these effects would be inconsequential on the timescales we consider. We use the hybrid symplectic and Bulirsch-Stoer (B-S) integrator built into Mercury6 \citep{m6}, and conserve energy to 1 part in $10^{8}$.

To perform these computationally intensive simulations, we make use of both the Open Science Grid \citep{osg1, osg2} accessed through the Extreme Science and Engineering Discovery Environment \citep[XSEDE;][]{xsede1}, and personal computational resources for the simulations used in this work, with the bulk of numerical integrations being run by the former. In all cases, each integration was run on a single core. At least 4000 realisations were run for each system. The integrations were generally completed in less than 24 hours, with most integrations taking less than 6 hours to run to completion. The simulations resulted in more than 3 Terabytes of data files, and took roughly 100,000 CPU hours to generate. We integrated all of the 15 Kepler systems with 5 or more planets. After all integrations were complete, we also ran integrations of three representative four-planet systems, in order to verify that they exhibit similar behaviour.

\begin{figure}[htbp] %  figure placement: here, top, bottom, or page
\centering
\includegraphics[width=5in]{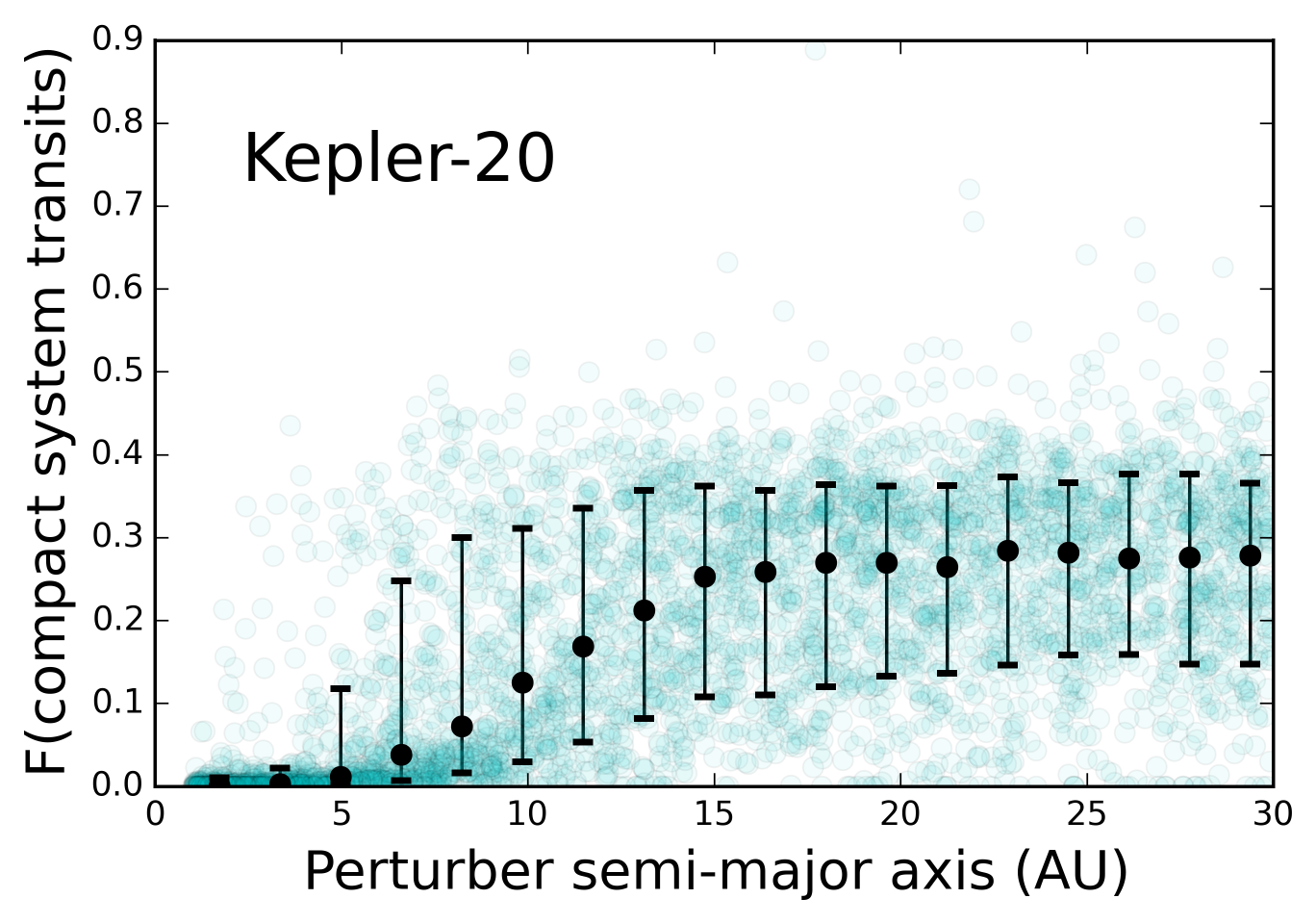}
\caption{The Kepler-20 system appears to be an outlier in our sample of multi-planet systems, and is highly susceptible to being perturbed into a CMT-unstable configuration by a companion. For any companion with orbital separation in the range $a=1-30$ AU, the inner system will be seen to be mutually transiting less than half (about one third) of the time. This result strongly suggests that if an additional close companion exists in this system, it is unlikely that we would see all of the planets in the inner system to be transiting.  }
\label{kep20}
\end{figure}

Figure \ref{colorbarred} and Figure \ref{kep20} show each individual realisation of Kepler-102 and Kepler-20, respectively, as circular points, where we plot the fraction of time a compact inner system is CMT-stable in the presence of a companion with the periastron value given on the $x$-axis. These plots (and all analyses presented in this paper) are marginalised over all of the other orbital elements chosen for the perturber. The population of points are significantly different for each system. Figures \ref{colorbarred} and Figure \ref{kep20} show two examples of systems: Kepler-20 is CMT-unstable a significant fraction of the time, for almost the full range of companion properties, although smaller orbital radii for the perturber do lead to an increase in  CMT-instability (as expected). On the other hand, Kepler-102 has a clear threshold at roughly 10 AU where, external to this point, companions generally do not disrupt the behaviour of the inner system.

\section{General Results: Limits on Unseen Companions in the Kepler Sample}
\label{aligned}
\label{sec:limits} 

In this section, we present the results of the simulations detailed in the previous section. 

\subsection{General Trends}

For all systems considered in this work, the general trend holds that pertubers closer to the compact system of planets causes a greater amount of movement out of the transiting plane by those inner planets. This mechanism can lead to the known planets in each system attaining non-transiting configurations (CMT-instability), meaning that they would not have been discoverable by the \emph{Kepler} mission. There are two modes of motion due to an exterior companion that can lead to the aforementioned effects: (1) the excitation of relative inclination angles between the planets in the inner compact system, leading to an increase in the width of the inclination plane containing the compact system planets, and (2) Kozai-esqe oscillations of the entire plane of transiting planets, which may precess together. Both of these effects may occur, and both are encapsulated in our numerical integrations. Mode (1) will determine whether or not a system of planets can be seen in transit from any line of sight. In contrast, mode (2) could lead to a situation in which the system is not seen in transit from Earth's line of sight, but could be seen from another line of sight. For the sake of definiteness, we define the term CMT-stability to mean that the planets are continually mutually transiting from the original line of sight observed by Kepler (more specifically, all of planets can be seen in transit more than $95$\% of the time). This definition thus implies that both oscillation modes (1) and (2), as defined above, lead to 
CMT-unstable systems.  

\begin{figure}[htbp] %  figure placement: here, top, bottom, or page
\centering
\includegraphics[width=5in]{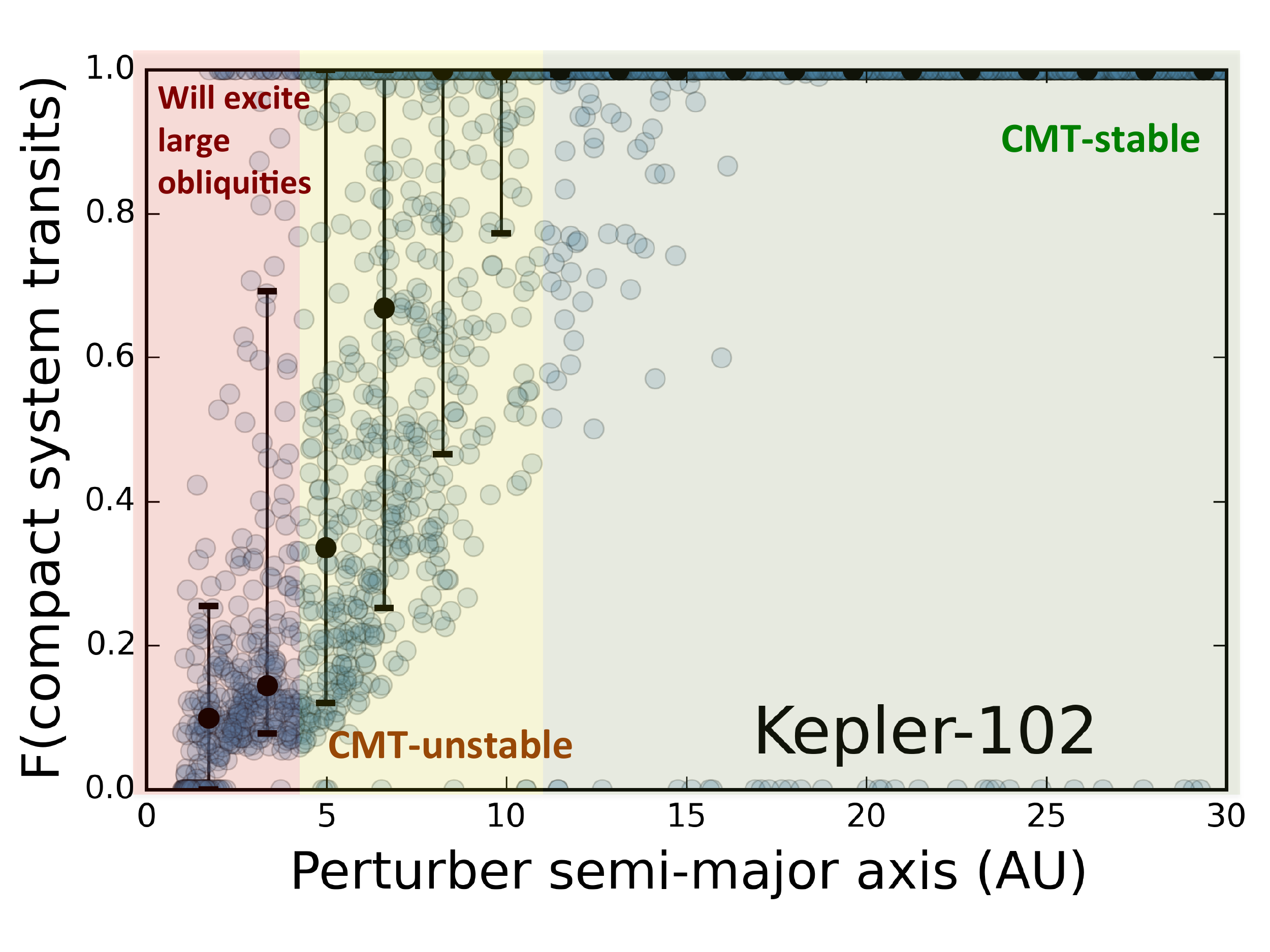} 
\caption{The fraction of the time that the entire compact system transits depends on the orbital elements of the perturbing body. Here, this fraction is plotted against periastron of perturber for Kepler-102, where the non-transiting disk prior was used to choose companion orbital elements. If the body is far away from the inner planets, then that companion exerts only minimal perturbations on the compact system of planets, and the system will be CMT-stable, just as expected if no perturber is present. For an intermediate range of perturber parameters, the system will be CMT-unstable, but no large obliquities will be excited. For a selection of perturbing bodies, not only will the inner system be CMT-unstable, but the orbits of the planets in the compact system will become highly misaligned relative to their initial locations. These three regions for Kepler-102 are shown and labelled in the figure. Similar plots can be made for all stars in our sample, and for each of the sets of priors. The location of the different regimes depend on the properties of the planets and their orbits.  }
   \label{colorbarred}
\end{figure}

Figure \ref{colorbarred} shows the Kepler-102 system, which serves as an example of the typical trend and demonstrates the different regimes of behaviour that can be excited by the injection of a perturbing companion into a known planetary system. 
In this figure (and everywhere in this paper), the criterion F(compact system transits) = 1 means that the inner system of planets (which includes only those discovered to be transiting in the Kepler data) is continually mutually transiting. In other words, all planets can be seen to be transiting from Earth's line of sight for all time in the presence of the any considered perturbing companion. In contrast, the criterion F(compact system transits) = 0 means that the inner compact system that was found to be transiting by the Kepler mission will never attain a mutually transiting configuration in the presence of that companion. Figure \ref{colorbarred} shows that for companions with orbits beyond 10 AU, the Kepler-102 system will continue to be mutually transiting. The fact that the Kepler spacecraft observed the Kepler-102 system to be transiting thus cannot exclude any companions beyond 10 AU. However, for companions with periastron less than $\sim5$ AU, the Kepler-102 compact system will attain a non-transiting configuration a large fraction of the time, making it less likely that we would have discovered it. 

From the data used to construct Figure \ref{colorbarred}, and the analogous plots for the other planetary systems, we have computed the minimum periastron distance that allows the inner systems to remain in a CMT-stable configuration 95\% of the time. This threshold at 95\% ensures that the systems are likely to remain observable by Kepler over secular timescales, but the exact value is arbitrary. These periastron values (computed for the 95\% threshold) are useful for comparing the relative stability of systems, and for predicting/constraining the possible locations for any additional massive, external companions in these systems. This analysis was carried out for each system in our sample using each of the three prior choices described in Section \ref{masses}. The results of this computation (for all systems and all priors) are presented in Table \ref{table:the_limits}.

As outlined above, these limits correspond to the companion periastron required for the Kepler compact system to remain CMT-stable 95\% of the time. The limits vary slightly between the three prior choices, demonstrating the effect of the priors of this dynamical analyses. The spread between the three prior choices $\delta p / p$ attains a median value of 11\% and a mean value of 13\% over the entire population of systems considered (the higher mean is due to the Kepler-32 system, which experienced particularly large variation between prior choices -- if Kepler-32 is excluded from the sample, then the mean and median $\delta p / p$ values become equal). As a result, for the systems considered in this work, the choice of priors affects our results for the threshold values of the companion periastron at the level of 10-15\%. 

Each of these systems show the same general trend: For companions with sufficiently large orbital separations, the inner system becomes effectively decoupled from the perturber, so that the system is expected to stay in a CMT-stable configuration. In this context, the definition of being ``sufficiently large'' is given by the periastron values listed in Table \ref{table:the_limits}. These results also depend on other properties of the compact systems. 
We provide limits in terms of periastron rather than semi-major axis, which folds in the distribution of eccentricity for each prior type. We also generated the limits in terms of semi-major axis, and the values were not significantly different. Periastron is a good variable to use here for two reasons: first, the periastron describes how close the outer planet gets to the inner system of planets, which controls the magnitude of mean motion-based perturbations; second, the high- and low-eccentricity cases (constructed and discussed in Section \ref{masses}) do not exhibit significantly different behaviour, so our parametrisation in terms of periastron will not occlude any physical effects. 
For each planetary system, Table \ref{table:the_limits} also lists the surface density of the system, the corresponding angular momentum, and the inner system size (given by the semi-major axis of the outermost planet of the inner system). The outer radius of the compact system is included because planets with larger semi-major axis have lower transit probabilities. Naively, it might seem that these systems would be easier to force into a non-transiting configuration. However, this expectation does not hold --- the systems with the largest inner system radius are not necessarily those which are least CMT-stable in the presence of a companion. The dynamics of the inner system, not just the value of $R_{*}/a$, determines the CMT-stability of a system. 

% Requires the booktabs if the memoir class is not being used
\begin{table}[htbp]
   \scriptsize
   %\topcaption{Table captions are better up top} % requires the topcapt package
   \begin{tabular}{@{} cccccccccc @{}} % Column formatting, @{} suppresses leading/trailing space
      \toprule
System &$N_{p}$   &Surface density &Angular momentum     &$a_n$ (AU)  &$p_{\rm{uniform}}$   &$p_{\rm{transit}}$  &$p_{\rm{disk}}$ & F$_{jup}$\\ 
 &  &&      &   &(AU)  & (AU) & (AU) & \\ 
      \cmidrule(r){1-9} % 
Kepler-102	& 5  &	136	$\pm$	26	&	20	$\pm$	4	&	0.17	&	8.15	&	7.03	&	6.97	& 2\%\\
Kepler-11	&  6 &	113	$\pm$	10	&	206	$\pm$	18	&	0.47	&	 $\ge$ 30$^{1}$  	&	 $\ge$ 30 	&	 $\ge$ 30 	& 0\%	\\
Kepler-122	& 5  &	406	$\pm$	51	&	131	$\pm$	16	&	0.23	&	8.57	&	8.3	&	7.2	& 0\%	\\
Kepler-169	&  5 &	55	$\pm$	5	&	54	$\pm$	5	&	0.36	&	14.7	&	13.15	&	11.7	& 4\%	\\
Kepler-186	&  5 &	19	$\pm$	1	&	16	$\pm$	1	&	0.43	&	19.5	&	16.4	&	15.2	& 1\%	\\
Kepler-20	& 5  &	77	$\pm$	7	&	65	$\pm$	8	&	0.35	&	 $\ge$ 30 	&	 $\ge$ 30 	&	 $\ge$ 30 		& 2\%\\
Kepler-292	& 5  &	534	$\pm$	61	&	54	$\pm$	7	&	0.14		&6.9	&	6.9	&	5.4	  	& 18\%\\
Kepler-296	& 5  &	98	$\pm$	14	&	37	$\pm$	6	&	0.26	&	11.6	&	10.1	&	10	  	& 1\%\\
Kepler-32	& 5  &	539	$\pm$	59	&	35	$\pm$	4	&	0.13	&	8.7	&	11.8	&	7.03  	& 0\%	\\
Kepler-33	&  5 &	473	$\pm$	39	&	272	$\pm$	23	&	0.25	&	 $\ge$ 30 	&	 $\ge$ 30 	&	 $\ge$ 30 	  	& 0\%\\
Kepler-444	& 5  &	48	$\pm$	3	&	1	$\pm$	1	&	0.08	&	6.3	&	5.5	&	5.1	  	& 8\%\\
Kepler-55	& 5  &	240	$\pm$	24	&	49	$\pm$	4	&	0.20		& 8.5	&	10.6	&	7.5	  	& 0\%\\
Kepler-62	& 5  &	11	$\pm$	2	&	48	$\pm$	9	&	0.72	&	 $\ge$ 30 	&	 $\ge$ 30 	&	 $\ge$ 30 	  	& 0\%\\
Kepler-84	&  5 &	184	$\pm$	21	&	81	$\pm$	9	&	0.25	&	7.4	&	9.8	&	7.2	  	& 2\%\\
Kepler-90	&  7 &	48	$\pm$	3	&	901	$\pm$	53	&	1.01		& $\ge$ 30 	&	 $\ge$ 30 	&	 $\ge$ 30 	  	& 0\% \\
      \cmidrule(r){1-9} % 
Kepler-150	&  4 &	352	$\pm$	29	&	75	$\pm$	6	&	0.19	&	8.3	&	6.9	&	5.6   	& 4\% \\
Kepler-197	& 4  &	82	$\pm$	4	&	13	$\pm$	1	&	0.16	&	6.9	&	6.9	&	6.7	  	& 11\%\\
Kepler-402	& 4  &	571	$\pm$	73	&	28	$\pm$	3	&	0.10		& 5.2	&	5.3	&	5.3	  	& 16\%\\
\bottomrule
\end{tabular}
\caption{Physical properties of each system considered in this paper, along with the derived limits for the periastron of possible perturbing companions, when $N_{p}$ is number of planets in each system. The periastron limit  $p_{\rm{\textbf{prior}}}$ is the value of perturber periastron above which the inner system is CMT-stable 95\% of the time, also marginalised over all other properties of the perturber (including periastron) for each selection of \textbf{prior}. The scale $a_n$ is the outer orbital radius of the inner, compact system. For the three different types of priors, slightly different limits are computed, with the transit prior generally producing the more dynamically quiet systems. See Section \ref{masses} for a description of each of the three priors used. A selection of four-planet systems, which were not part of the original sample but included to check if the trends shown in Figure \ref{compare_fig} appear to persist for additional systems, are presented in the lower part of the table. $\delta p / p$ between the three priors is a median of 11\% and a mean of 13\% for all systems, demonstrating that prior choice can affect dynamical stability in analysis of this nature to the 10-15\% level. The final column, F$_{jup}$, is the results of a separate experiment, where Jupiter-type planets (1 M$_{jup}$ at 5 AU, with inclination and eccentricity similar to Jupiter's values) were injected into the system instead of the previously discussed perturbers. F$_{jup}$ is the percentage of these realisations that were CMT-stable in the presence of this true Jupiter-like planet. (1) 30 AU is the maximum orbital separation tested for the perturber in our simulations, so this designation implies that the presence of any perturber within 30 AU causes the inner system to be CMT-unstable according to our simulations. The true limit for CMT-stability 95\% of the time cannot be determined from our simulations.   }  
\label{table:the_limits}

\end{table}

\subsection{Surface Density as a predictor for susceptibility to perturbations }

\begin{figure}[htbp] %  figure placement: here, top, bottom, or page
\centering
\includegraphics[width=5in]{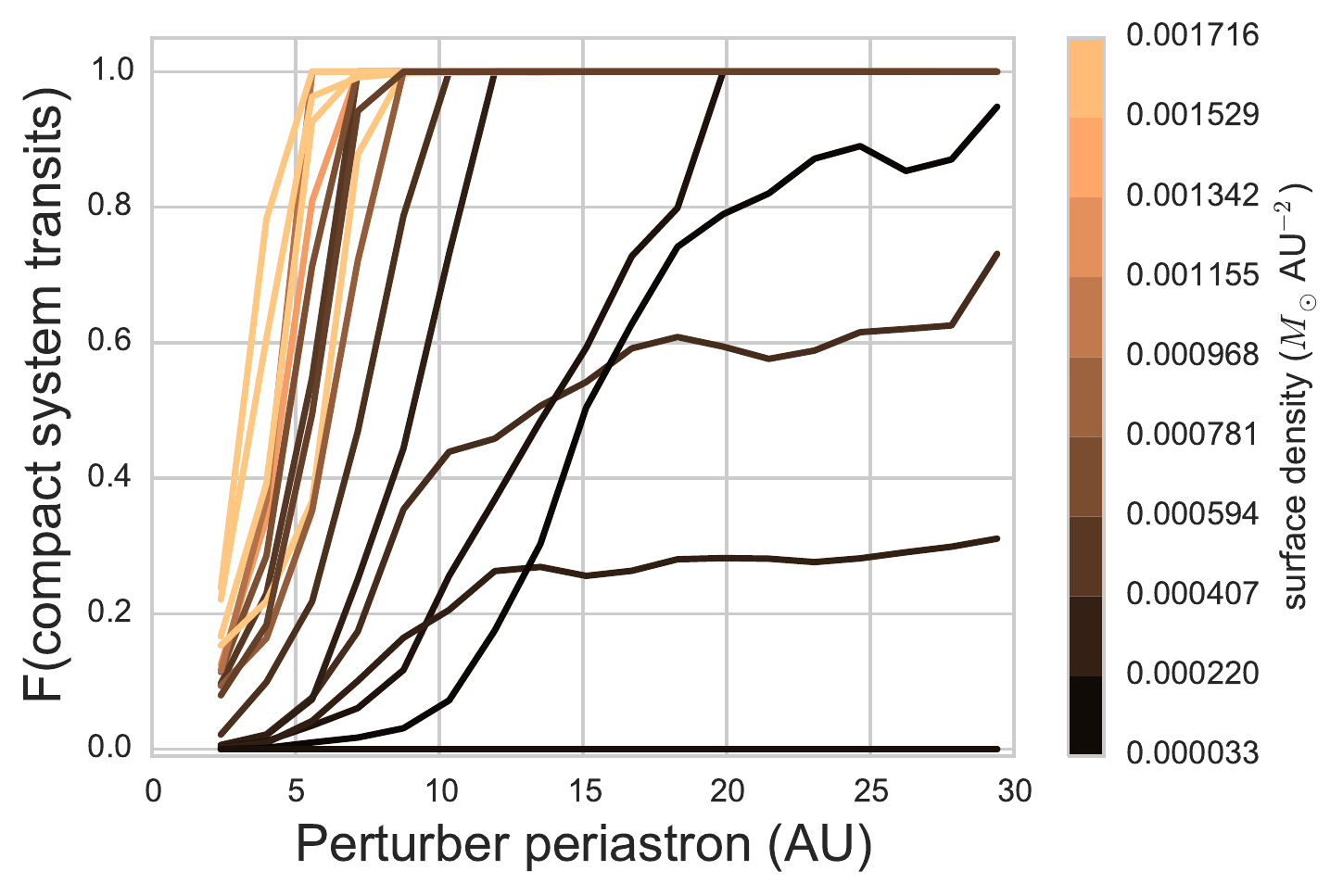} 
\caption{ The fraction of time a system is CMT-stable as a function of the periastron of the injected perturbing body, marginalised over all other properties of the perturber. Each line represents a different system, with the colour of the line showing the surface density of the compact inner system (computed using Equation \ref{fig:sd_fig}). The colour bar on the right shows the scale. Systems with higher surface densities tend, with good uniformity, to allow a larger array of perturbing companions without becoming CMT-unstable. }
\label{compare_fig}
\end{figure}

The results presented in Table \ref{table:the_limits} give the relative radii within with you expect significant misalignment to arise. Naively, it might seem that non-transiting planets will be observed more often in systems where the inner system has a larger outer radii (i.e., the semi-major axis of the furthest-out planet is larger). However, this value $a_{out}$ does not predict the susceptibility of an inner system to perturbations. Instead, a better tracer of how susceptible a system is to perturbations is the surface density of the inner system of planets. 
In Figure \ref{compare_fig}, we plot the fraction of time that a system is CMT-stable as a function of the periastron of the distant perturbing planet. The lines are colour-coded by the surface density of the observed compact inner systems. The plotted surface densities were calculated by computing the surface density, $\Sigma$, of each realisation of the system according to the definition 
\be
\Sigma = {1 \over \pi (a_{n}^2 - a_{1}^2)}
\sum_{i=1}^{n} m_{i} \,, 
\label{fig:sd_fig}
\ee
where $n$ is the number of planets in the system, with planet $n$ being the outermost planet, $m_i$ are the planetary masses, and $a_i$ are the orbital radii. Since we draw masses and periods from observationally-inspired priors, the exact value of the surface density varies between the different trials of our numerical integrations. For this plot, the chosen value of surface density was taken to be the median value of the surface density for all realisations of that system. The error on surface density, which is given for each system in Table \ref{table:the_limits}, is given by the 1$\sigma$ spread over all of the realisations.

If the surface density is higher, then the inner system acts more like a single ring, and perturbing bodies need to be closer to the inner system in order to excite individual inclinations away from a mutually transiting configuration. It is clear from Figure \ref{compare_fig} that surface density maps (almost) monotonically onto the periastron distance at which CMT-instability does occur. On the other hand, the periastron of the perturbing body is not the only quantity of interest. To construct Figure \ref{compare_fig}, we marginalised over the orbital parameters of the outer companion, using the non-transiting disk prior (see Section \ref{masses}). An analogous plot can be constructed for the other priors types, but all three priors will produce results consistent on the 10-15\% level (the value of periastron for the companion required to render the systems CMT-unstable varies by this amount over the different choices of priors). 

Intuitively, the impact periastron distance has on the CMT-stability of the inner system makes sense because the periastron distance controls how closely the perturbing body passes to the inner system. In \citet{paper1}, we used secular theory to evaluate the long-term behaviours of these same systems. The expanded disturbing function in secular theory depends most strongly on semi-major axis, so the dynamics considered here must depend sensitively on $a$. In addition, eccentricity allows the companions to pass closer to the inner system. Previous studies have shown that the periastron of the perturbing companion is the most important variable for ejecting planets \citep{evadavid} and for stifling the formation of planets in binary systems \citep{quintana}. As a result (and for brevity), we do not provide plots for CMT-stability as a function of mass, eccentricity, inclination, and other orbital properties.
%, although such plots can be constructed and %are available upon request. 

\begin{figure}[htbp] %  figure placement: here, top, bottom, or page
\centering
\includegraphics[width=5in]{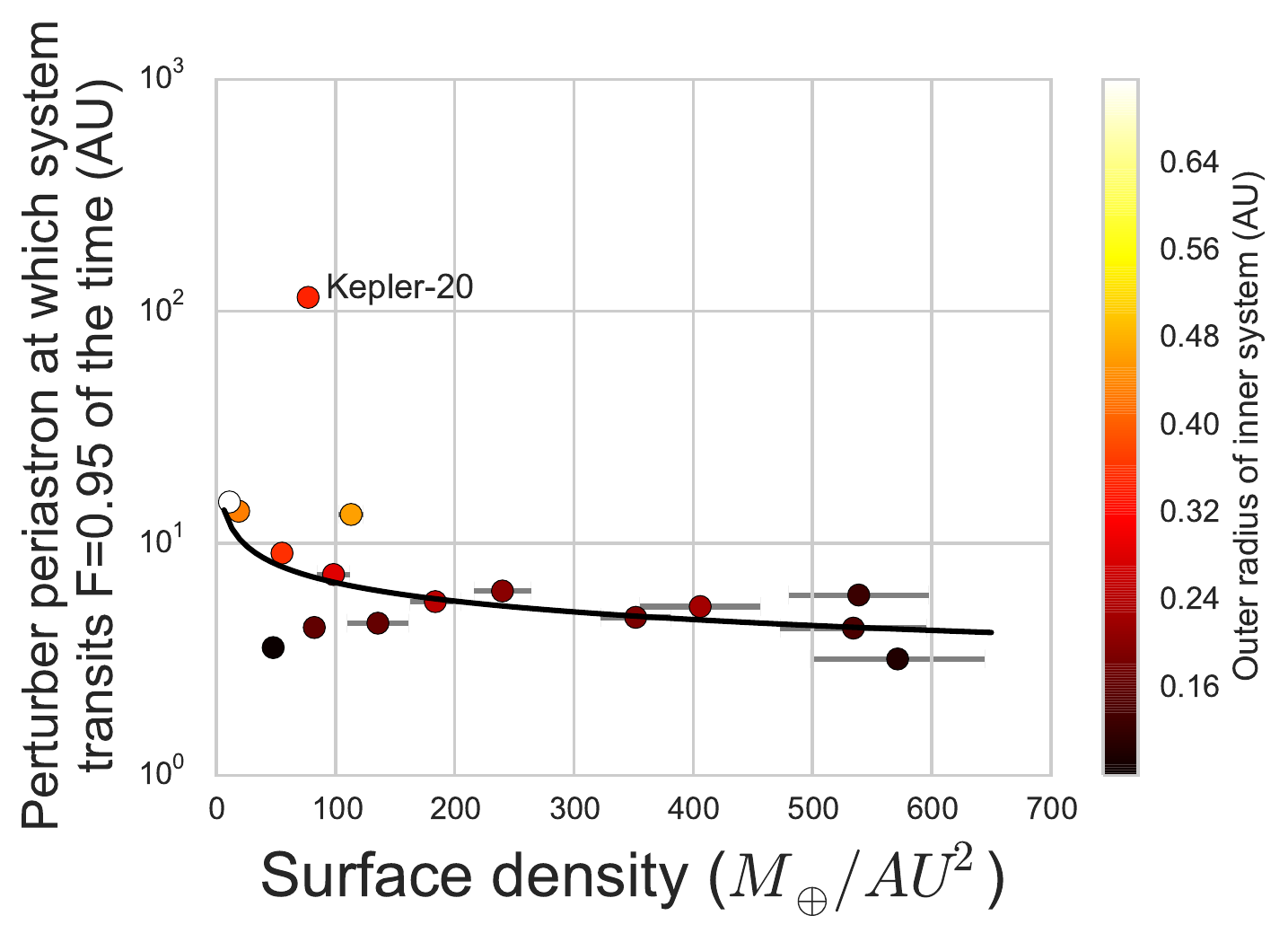} \caption{The periastron of the most distant perturbing planet that allows the inner compact system to remain CMT-stable (mutually transiting 95\% of the time), shown here as a function of the surface density of the observed compact inner systems. When the perturbing body has a sufficiently large orbit, it becomes effectively decoupled from the inner system. It is important to note that the lower surface density systems tend to have a larger radius within which companions would lead the inner system to be CMT-unstable. Kepler-20 appears to be a clear outlier compared to the other systems, with a large range of potential companions leading the inner system to be CMT-unstable. If the error bar is not visible, the error is smaller than size of the plotting symbol. The curve shows a model fit to all points except that of Kepler-20.  }
\label{surf_dens_fig}
\end{figure}

In Figure \ref{surf_dens_fig}, we plot the periastron of the most distant perturbing planet that allows the inner compact system to remain CMT-stable (mutually transiting 95\% of the time) as a function of the surface density of the observed compact inner systems. The periastron values were taken from interpolated versions of the curves shown in Figure \ref{compare_fig}. For most of the transiting systems in our sample, a perturbing companion must be roughly 10 -- 20 AU away (or more) from the host star in order for the compact inner system to have a high probability of being observed in the configuration discovered by the Kepler mission. Note that systems with the largest orbital radius of their respective inner systems tend to have the lowest surface densities. This trend could suggest that they are either stalled at an earlier stage in their migratory histories (having not collapsed to the size of the smallest, densest multi-planet systems) or that our observations are incomplete. It remains possible that this trend reflects the diversity of possible planetary configurations.

Using the limited number of systems analysed in this work, we can derive an approximate scaling relation that delineates how far a perturber must reside from the central body in order to excite CMT-instabilities in a transiting system with a given surface density. The model curve plotted in Figure \ref{surf_dens_fig} is a power-law, which was fit with a simple least-squares optimisation, with approximate best-fit functional form $p = 6.60\ {\rm AU} \, \left( {\Sigma / \Sigma_0}\right)^{-0.27} \,,$ where $\Sigma$ is the surface density of the inner system of planets and $p$ is the periastron beyond which companions do not disrupt the CMT-stability of the system 95\% of the time. The constant $\Sigma_0$ = 100 $M_\oplus$ AU$^{-2}$ is a reference surface density. 
%This rough scaling relation can be used to predict the radius outside of which companions to these tightly-packed multi-planet systems would not have a significant effect on the CMT-stability on the inner system. 
This relation defines the region outside of which a companion could exist without affecting the inner system, i.e., so that it would not alter the transits of the observed planets. In contrast, any companions found within this boundary could cause significant misalignment of the orbital inclinations, perhaps knocking some inner planets out of an observable, transiting, configuration. 

This relation is limited in two ways: First, the number of systems under consideration is small, by necessity, so that additional systems could display more complicated behaviour. Second, the relation is approximate, and depends on both the priors used and the relatively few low surface-density systems. As a result, this relation does not provide a definitive statement on the companion status of any particular system.  

\subsection{Examining the effect of Jupiter}

Our own solar system has a gas giant planet with mass $m_P=1 M_{\rm jup}$ with semi-major axis $a\approx5$ AU. The results of Table \ref{table:the_limits} indicate that 5 AU often falls within the radius at which additional companion cause the inner system to become misaligned. As a result, a true Jupiter analogue is unlikely to exist in any of these systems. However, the previous trials do not directly test for this possibility because the inclination variation of all three prior choices is larger than that of a true Jupiter analogue. For this reason, we performed an additional set of numerical simulations. In the work described thus far, we examined the effect that a perturber of varying orbital properties could have on compact systems of planets. However, it is also interesting to determine how the presence of a true Jupiter analogue would affect the CMT-stability of these systems. Toward this end, we ran another 100 realisations per system, including a Jupiter analogue as the perturber with the following priors on its orbital elements: The mass is taken to be 1 $M_{\rm jup}$ and the semi-major axis $a$ = 5 AU. The eccentricity is drawn from a uniform distribution in the range [0,0.05], and the inclination is drawn from a uniform distribution with a full width of 6 degrees, thus allowing the planet to attain a maximum inclination of three degrees out of the plane containing the inner system. All of the other orbital angles were randomised. We integrated these new realisations for 10 Myr and evaluated the fraction of the realisations that remained CMT-stable (mutually transiting 95\% of the time). These percentages are reported in the final column of Table \ref{table:the_limits}. In all but two of the systems containing more than four planets, more than 95\% of trials were CMT-unstable. All of the systems were CMT-unstable a majority of the time. This finding indicates two things: First, it is unlikely that a Jupiter analog planet exists in any of these systems. Second, if such a system were to host a Jovian analog planet, it would (generally) lead to oscillations of the inclination angles of the orbits of the inner system. As a result, there could be additional unseen planets in the compact, inner part of that system. 

The discussion thus far has not taken into account the stellar-spin axis, in particular its direction with respect to the orbital angular momentum vectors of the compact inner system. 
\citet{tim} found that the obliquities of multi-planet systems tend to be lower than those in single-planet systems. \footnote{Some exceptions are expected. As one example, the Kepler-108 system, which hosts two planets, is thought to be misaligned \citep{misalign108}. Another example is the well-studied, misaligned Kepler-56 system \citep{huber}.}
%In addition, \citet{temp6250} found that cool stars (those with surface temperature below 6250 K) also tend to have low obliquities. Since all of the stars in our sample have $T_{\rm eff}<6250$ K \citep{systems2010, systems2010b, systems2013, systems2013b, systems2014b} and all are multi-planet systems, we expect the majority of them to have low obliquity.
If the stellar-spin is observed to be aligned with the orbits, as seems to be common from observational results for multi-planet systems, then the most likely scenario is that the all of the angular momentum vectors point in their original directions.  It is unlikely that both the stellar spin and all of the planetary orbits were disrupted in such a way that they maintain alignment. In contrast, systems with an observed misalignment between the stellar spin-axis and the orbital angular momentum vectors could have a variety of dynamical histories. One possibility is that the orbital inclination angles for the entire inner compact system are oscillating as a whole in response to a perturbing companion. \citet{chelsea} and \citet{moveplane} both consider this mechanism in depth, and find that it is a feasible method of causing planet-star misalignment. 
%The work presented here indicates that such coherent oscillations rarely occur. 
Moreover, a number of other effects could lead to the stellar spin pointing in a different direction than the angular momentum of the planetary system. Possible processes include natal misalignment, interactions with unbound bodies, tidal precession, and many others. All of these effects should be explored in future work.

\section{Results for Specific Systems}
\label{sec:insights}

In addition to the general results presented in Section \ref{sec:limits}, this analysis also produces predictions and insights for individual systems. Here we consider the cases of Kepler-20 and WASP-47.

\subsection{Kepler-20}

Kepler-20 appears to be an outlier among the systems considered in this work, as shown by its unique placement in Figure \ref{surf_dens_fig}. Kepler-20 requires an unusually large orbital separation between the added perturber and the inner system in order for the inner system to remain CMT-stable. Taken alone, this placement on the plot indicates that Kepler-20 is particularly susceptible to the effect of a companion; it is quite easy for a perturbing companion to knock the inner system into a non-transiting configuration.

This seems to be evidence of one of two things: either (1) there is no external companion in this system, because its existence would not allow the entire inner system to transit, or (2) the entire inner system is not transiting (the system is NOT continually, mutually transiting) and there is another planet that we do not know about.

The Kepler-20 considered in this work only included five planets, the original five reported in \citet{kep20ecc} and \citet{kep20-1}. After we had completed our simulations for this work, however, this system was found to host an additional planet with minimum mass 20 $M_{earth}$ in an approximately 34-day orbit. This places the orbit of the new planet \emph{between} the 20-day period Kepler-20f and the 78-day Kepler-20d \citep{twittertalk}. Because this newly detected planet lies between the orbits of the previously known planets, and because it was not observable by Kepler but is observable in the RV, it is likely to be slightly out of the transiting plane. The Kepler-20 system thus hosts five transiting planets and one non-transiting planet, with all six planets packed in a compact system. We now know that (2) is the correct conclusion for this particular system, which does not exclude the possibility that there is also an exterior, perturbing companion. 

Additional work will help determine if such a companion is present, or if another explanation exists for the unusual configuration of Kepler-20. In general, compact planetary systems tend to have regularly spaced orbits, where all of the planets are seen in transit. If a system has a gap in its orbital spacing, there could be an additional massive planet of the type considered here. 

%The relative stability of Kepler-20, compared to the other systems in our sample, suggests the following: Although the other systems in our sample are significantly less likely to have unseen, short- or long-period companions, it may be possible. An analysis of the type presented in this work could inform which planets are most likely to host extra planets, and thus which are likely to be most fruitful RV follow-up targets.

\subsection{WASP-47}
\label{WASP47} % citations XXX

The WASP-47 system was known to host a hot Jupiter \citep{wasp47disc}, was later found  to contine two additional planetary companions with orbital periods less than 10 days \citep{wasp47}. A super-Earth companion was discovered just inside the orbit of the hot Jupiter and a Neptune-sized planet was found just outside. \cite{marion} reported simultaneously that the system also has a Jovian external perturber, a companion with $m \sin{i}$ = 1.24 $M_{\rm jup}$ and a period of 572 days. The hot Jupiter in this four-planet system was also found \citep{roberto} to have its orbital angular momentum vector aligned with the stellar spin axis of the star (implying that the two other transiting planets are also roughly aligned). 

The fact that the inner three planets in the WASP-47 system appear to remain in their birth plane suggests that the external companion must allow for the persistent CMT-stability of the inner three-planet system. Because the RV measurements of the outer companion only determine the quantity $m \sin{i}$, and not the true mass, we do not know whether the companion is a highly inclined brown dwarf or a roughly co-planar Jovian planet. Using the techniques from this paper, however, we can place probabilistic limits on the inclination and mass of the outer body in this system. 

\begin{figure}[htbp] %  figure placement: here, top, bottom, or page
\centering
\includegraphics[width=3.7in]{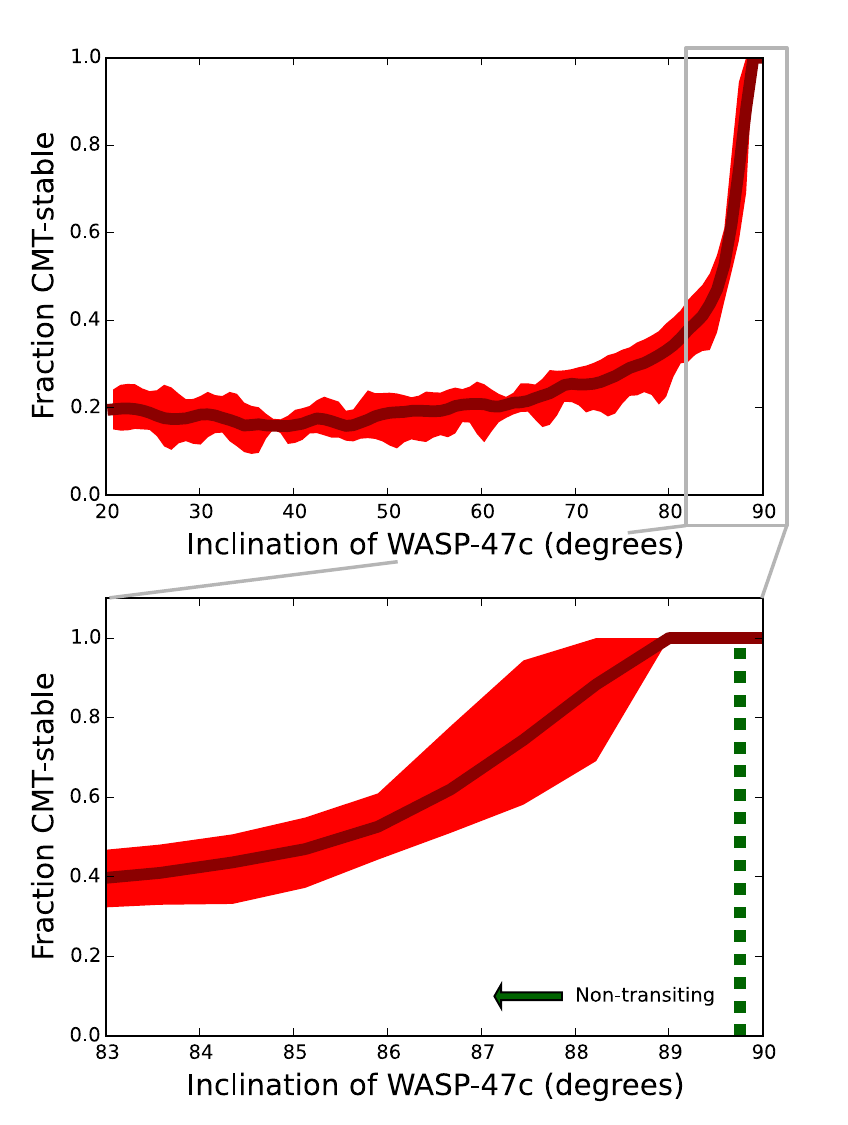} 
\caption{WASP-47, a good example of a planetary system with both a tightly packed inner system (WASP-47e, -47b, and -47d) and an outer perturber (WASP-47c) is a good test case for the methods used in this paper. The line shows the median fraction of time the inner system is CMT-stable for a given inclination, and the color-block shows the entire parameter space populated by the 1000 trials. The CMT-stability of the inner system decreases to a paltry 20\% if the outer companion is allowed to attain a significant inclination out of the plane containing the other planets (which, in parameterization, is the plane centered at 90 degrees). There is a large area in parameter space that allows the inner system to be CMT-stable and the outer planet to be non-transiting (see lower panel). }
\label{wasp47fig}
\end{figure}

Toward that end, we ran 1000 integrations of the WASP-47 system for 10 Myr each. The orbital properties of the inner three planets were drawn from the posteriors found in \citet{wasp47} from the transit and transit-timing-variation fits. For completeness we note that \citet{fei} also provides mass estimates of the three inner planets from RV measurements, but the results are consistent with the TTV-estimated masses used here (within 1$\sigma$ uncertainties). The inclination of the outer planet was allowed to vary over the full range from 0 to 90 degrees (we expect the 90 degree to 180 degree range to be symmetric; notice also that $i$ = 90 degrees is defined as the midpoint of the transiting plane, i.e., where the inner three planets reside). The mass of the outer planet was chosen for each trial to satisfy the observed $m \sin{i}$ measurement within the reported errors given the assigned inclination for that trial. The results are plotted in Figure \ref{wasp47fig}.

In this systems, the spin-axis of the star is observed to be aligned with the orbital angular momentum of the inner three planets, so that the system is likely to have formed with such an alignment. The inclination of the orbit of the outer companion is unknown {\it a priori}. Our numerical results show that in order for the inner system to remain CMT-stable most of the time, the inclination of WASP-47c cannot be more than 2-3 degrees out of the transiting plane at 90 degrees (the plane that contains the inner three planets). Since the semi-major axis of WASP-47c is large, the inclination of its orbit can easily be large enough that the planet does not transit (as observed) but still lies within a couple degrees of the plane (as required by CMT-stability) --- the inclination only needs to be a few tenths of a degree to keep WASP-47c from transiting. Notice also that further refinement of the masses of the inner system of planets will allow for a more robust exploration of this constraint. 

The observed alignment between the spin-axis of the star and the angular momentum of WASP-47b's orbit, combined with the suggestion that WASP-47c is likely roughly coplanar with all three planets of the inner system, indicates a dynamically quiet history for the system. Indeed, it is easier to knock a system out of alignment than return it to alignment, so it is unlikely (although not impossible) that the WASP-47 system is dynamically active and we see it at an opportune moment. It is more likely that WASP-47 formed and migrated (in either order) in a dynamically quiet manner.

% explain what could cause misalignment 
% show how results of previous section change if you allow spin-orbit obliquities to vary

%make some plots synthesiing all the results from the sims / analytical theory -> did that, there's like 10 of them
% how can you predict the stability of a system by just looking at the planet properties? surface denity or angular momentum plots here -donsies
% If more planets are discovered in these systems, limits get (stronger?) -maybe kepler 20 deserves its own section

\section{Conclusions}
\label{sec:conclude} 

In this work, we have considered a collection of 18 Kepler multi-planet systems and evaluated their CMT-stability in the presence of a perturbing Jovian companion with a semi-major axis between 1 and 30 AU. 
A system that we define to be \textbf{C}ontinually \textbf{M}utually \textbf{T}ransiting is expected to remain transiting from our line of sight over many secular periods. In contrast, a system in a CMT-unstable configuration would be not be visible in transit all the time --- sometimes one subset of planets might be visible, and another time a different subset might be visible. The sample we consider was chosen to include all observed systems with five or more planets, along with some additional four-planet systems for comparison. The total number of systems analyzed was limited by  computational resources.

%We would have liked to test all observed Kepler systems, and such an endeavor (perhaps eventually with K2 or TESS planets) would likely be fruitful in the future if computational limits can be overcome. 

This work presents both general results and specific results for particular systems. We find that massive, close-in companions to the observed Kepler multi-planet systems will lead to CMT-instability. Table \ref{table:the_limits} gives the limits for each of the systems analyzed in this work. For most of the systems in the sample, the fraction of time that the system remains CMT-stable approaches unity at a well-defined value of the companion periastron (see Figure 5). Moreover, these periastron values fall in the range $p=5-15$ AU for the majority of the systems. The specific values, for each of the sets of priors, are given in Table \ref{table:the_limits}. 

From the population of systems explored here, we can also deduce general limits on the presence of possible companions. These systems can remain CMT-stable provided that the companion periastron is greater than $p=5-30$ AU, although some systems saturate this outer limit so that any companions must reside in even wider orbits. At one extreme, systems can be rendered CMT-unstable with a planet comparable in mass and orbital radius to Jupiter (e.g., Kepler-402). Most systems require companions to reside beyond $p\sim10$ AU, whereas some systems are so sensitive to inclination angle oscillations that companions must lie beyond 30 AU. We can thus draw the following conclusion: If exterior companions (with periastron inside of 10 AU and masses of a few $M_{\rm jup}$) were common, we would often expect to see significant mis-alignments in at least some of the observed multi-planet systems. Since we do not see this effect in the sample, it is unlikely that this type of companion is common in the observed multi-planet systems.

We also find that the surface density of a compact system of planets can serve as a good prediction for whether systems are CMT-stable. Low surface density systems tend to be more susceptible to perturbations by additional companions (see Figure 6). 

In addition to the general analysis summarized above, we considered the effects of adding Jupiter-analog planets (with mass of 1 $M_{\rm jup}$ and $a$ = 5 AU) into the compact systems. The CMT-stability of the resulting systems is greatly compromised; these results are also presented in Table \ref{table:the_limits}. This numerical experiment shows that none of the compact Kepler systems are allowed to have a Jupiter analog. If such a planet were present, the system would (almost always) be driven to CMT-instability, so that the full set of planets would not be observed in transit.

This dynamical treatment also provides results for particular systems, specifically, WASP-47 and Kepler-20. The WASP-47 system is one where an analysis of this nature proves particularly useful: In order for the inner three planets of the system to remain CMT-stable, the inclination angle of the fourth (more distant) companion must be small, which in turn implies that it has a planetary mass (rather than a larger mass with the orbit observed at high inclination). In other words, CMT-stability of the WASP-47 system predicts that the fourth planet must be at low inclination with mass $m \approx m \sin i$. The finding that all four planets in this system exist in a roughly coplanar configuration has implications for the formation scenario of this system, and suggests that the system formed and migrated dynamically quietly. More work should be done to understand the dynamical history of this particular system, whose inclination spread is analogous to our own solar system.

Kepler-20 is another intriguing system. Our simulation results for the five-planet Kepler-20 showed that the system was particularly susceptible to excitations from a companion at a large range of orbital separations. Kepler-20 became CMT-unstable a large fraction of the time. A solution was recently found for this puzzling observation: Kepler-20 was recently discovered to have a sixth planet orbiting in a non-transiting configuration in between the orbits of the previously discovered planets. Since not all planets in the inner system of Kepler-20 actually transit, this case is actually a CMT-unstable system, as the inner compact system of planets does not continually, mutually transit. 

In this work, we considered three choices for the priors used to specify the orbital properties of the companion: a uniform prior, a transit-inspired prior, and non-transiting disk prior. The results described above are largely insensitive to the choice of priors. We can quantify this effect as follows. The main result of this analysis is the threshold value of companion periastron, where companions must have larger values in order to not render the inner system CMT-unstable. These values are listed in Table \ref{table:the_limits}. 
The spread between the three prior choices $\delta p / p$ are on the order of 10\%-15\%, so that this variation provides an estimate on the uncertainty of our quoted periastron thresholds. 

For all three choices of priors, we have identified some Kepler systems as being probabilistically CMT-unstable in the presence of any additional perturbing companion (of the type considered here). This result implies that if any additional, perturbing body were in the system, we would not expect all of the inner system planets to transit (at least not most of the time). These highly susceptible systems are Kepler-11, Kepler-20, Kepler-33, Kepler-62, and Kepler-90. If these systems actually have the planetary properties that are currently reported, then these systems are unlikely to host additional companions. In other words, either these planetary systems have no additional companions (of the class considered here) or their properties are not determined correctly.

It is interesting to note that the systems least capable of hosting an additional companion (see Table \ref{table:the_limits}) fall into two categories. First, systems may have a large surface density (Kepler-33), which may be too tightly packed, such that any small perturbation from a companion excites the system into a non-CMT-stable state. Second, and more common, are systems like Kepler-90, Kepler-62, Kepler-20, Kepler-186, and Kepler-11, which have lower surface densities. The planets in these systems will attain higher inclinations more easily.
The results of this work imply that the second case is more common. In other words, planetary systems with surface densities low enough to be easily perturbed are more common than the extremely dense ones. An extended analysis of additional Kepler, K2, and TESS systems will determine if this apparent trend holds. 
As more systems are discovered and characterized, the techniques of this paper will be useful in constraining their possible architectures.

\bigskip
\noindent
{\bf Acknowledgments:} This work used the Extreme Science and Engineering Discovery Environment (XSEDE), which is supported by National Science Foundation grant number ACI-1053575. This research was done using resources provided by the Open Science Grid, which is supported by the National Science Foundation and the U.S. Department of Energy's Office of Science. JCB is supported by an NSF graduate fellowship. We would also like to thank Daniel Tamayo, Christopher Spalding, and Andrew Vanderburg for useful conversations. Finally, we thank the referee for many useful comments that improved the manuscript.


\begin{thebibliography}{99} 

\bibitem[Adams et al.(2013)]{ao13} Adams, E.~R., Dupree, A.~K., Kulesa, C., \& McCarthy, D.\ 2013, \aj, 146, 9 

\bibitem[Ballard \& Johnson(2016)]{ballard14} Ballard, S., \& Johnson, J.~A.\ 2016, \apj, 816, 66 

\bibitem[Baranec et al.(2016)]{ao162} Baranec, C., Ziegler, C., Law, N.~M., et al.\ 2016, \aj, 152, 18 

\bibitem[Barclay et al.(2015)]{barclay} Barclay, T., Quintana, E.~V., Adams, F.~C., et al.\ 2015, \apj, 809, 7 

\bibitem[Batalha et al.(2013)]{systems2013} Batalha, N.~M., Rowe, J.~F., Bryson, S.~T., et al.\ 2013, \apjs, 204, 24 

\bibitem[Batygin et al.(2011)]{secular1} Batygin, K., Morbidelli, A., \& Tsiganis, K.\ 2011, \aap, 533, A7 

\bibitem[Batygin 
\& Adams(2013)]{2013ApJ...778..169B} Batygin, K., \& Adams, F.~C.\ 2013, \apj, 778, 169 

\bibitem[Batygin \& Brown(2016)]{planet9} Batygin, K., \& Brown, M.~E.\ 2016, \aj, 151, 22 

\bibitem[Becker et al.(2015)]{wasp47} Becker, J.~C., Vanderburg, A., Adams, F.~C., Rappaport, S.~A., \& Schwengeler, H.~M.\ 2015, \apjl, 812, L18 

\bibitem[Becker 
\& Adams(2016)]{paper1} Becker, J.~C., \& Adams, F.~C.\ 2016, \mnras, 455, 2980 

\bibitem[Brakensiek \& Ragozzine(2016)]{corbits} Brakensiek, J., \& Ragozzine, D.\ 2016, \apj, 821, 47 

\bibitem[Borucki et al.(2010)]{systems2010} Borucki, W.~J., Koch, D., Basri, G., et al.\ 2010, Science, 327, 977 

\bibitem[Borucki et al.(2011)]{systems2011} Borucki, W.~J., Koch, D.~G., Basri, G., et al.\ 2011, \apj, 736, 19 

\bibitem[Borucki et al.(2013)]{systems2013b} Borucki, W.~J., Agol, E., Fressin, F., et al.\ 2013, Science, 340, 587 

\bibitem[Buchhave et al.(2016)]{twittertalk} Buchhave, L.~A., Dressing, C.~D., Dumusque, X., et al.\ 2016, arXiv:1608.06836 

\bibitem[Burke et al.(2015)]{burke} Burke, C.~J., Christiansen, J.~L., Mullally, F., et al.\ 2015, \apj, 809, 8 


\bibitem[Carrera et al.(2016)]{carrera} Carrera, D., Davies, M.~B., \& Johansen, A.\ 2016, \mnras, 463, 3226 

\bibitem[Caselli et al.(2002)]{cores} Caselli, P., Benson, P.~J., Myers, P.~C., \& Tafalla, M.\ 2002, \apj, 572, 238 


\bibitem[Chambers(1999)]{m6} Chambers, J.~E.\ 1999, \mnras, 304, 793 

\bibitem[Dai et al.(2015)]{fei} Dai, F., Winn, J.~N., Arriagada, P., et al.\ 2015, \apjl, 813, L9 


\bibitem[David et al.(2003)]{evadavid}
David, E.-M., Quintana, E. V., Fatuzzo, M., \& Adams, F. C. 2003, PASP, 115, 825

\bibitem[Dressing \& Charbonneau(2013)]{dressing} Dressing, C.~D., \& Charbonneau, D.\ 2013, \apj, 767, 95 

\bibitem[Fabrycky 
\& Winn(2009)]{2009ApJ...696.1230F} Fabrycky, D.~C., \& Winn, J.~N.\ 2009, \apj, 696, 1230 

%\bibitem[Fabrycky et al.(2012)]{kep32mass} Fabrycky, D.~C., Ford, 
%E.~B., Steffen, J.~H., et al.\ 2012, \apj, 750, 114 

\bibitem[Fabrycky et al.(2012)]{systems2012} Fabrycky, D.~C., Ford, E.~B., Steffen, J.~H., et al.\ 2012, \apj, 750, 114 


\bibitem[Fabrycky et al.(2014)]{hillstab1} Fabrycky, D.~C., 
Lissauer, J.~J., Ragozzine, D., et al.\ 2014, \apj, 790, 146 

\bibitem[Fang 
\& Margot(2012)]{fangmar} Fang, J., \& Margot, J.-L.\ 2012, \apj, 761, 92

\bibitem[Fielding et al.(2015)]{cob1} Fielding, D.~B., McKee, C.~F., Socrates, A., Cunningham, A.~J., \& Klein, R.~I.\ 2015, \mnras, 450, 3306 

\bibitem[Foreman-Mackey et al.(2016)]{longperiod} Foreman-Mackey, D., Morton, T.~D., Hogg, D.~W., Agol, E., \& Sch{\"o}lkopf, B.\ 2016, \aj, 152, 206 

\bibitem[Fressin et al.(2012)]{kep20-1} Fressin, F., Torres, G., Rowe, J.~F., et al.\ 2012, \nat, 482, 195 

\bibitem[Fressin et al.(2013)]{fressin} Fressin, F., Torres, G., Charbonneau, D., et al.\ 2013, \apj, 766, 81 

\bibitem[Gautier et al.(2012)]{kep20ecc} Gautier, T.~N., III, 
Charbonneau, D., Rowe, J.~F., et al.\ 2012, \apj, 749, 15 

\bibitem[Gratia \& Fabrycky(2017)]{moveplane} Gratia, P., \& Fabrycky, D.\ 2017, \mnras, 464, 1709 


\bibitem[Goodman et al.(1993)]{goodman} Goodman, A.~A., Benson, P.~J., Fuller, G.~A., \& Myers, P.~C.\ 1993, \apj, 406, 528 

\bibitem[Hansen(2016)]{priors_hansen} Hansen, B.~M.~S.\ 2016, arXiv:1608.06300 

\bibitem[Hellier et al.(2012)]{wasp47disc} Hellier, C., Anderson, D.~R., Collier Cameron, A., et al.\ 2012, \mnras, 426, 739 

\bibitem[Howard et al.(2012)]{howard} Howard, A.~W., Marcy, G.~W., Bryson, S.~T., et al.\ 2012, \apjs, 201, 15 

\bibitem[Huang et al.(2016)]{chelsea} Huang, C.~X., Petrovich, C., \& Deibert, E.\ 2016, arXiv:1609.08110 

\bibitem[Huber et al.(2013)]{huber} Huber, D., Carter, J.~A., Barbieri, M., et al.\ 2013, Science, 342, 331 


\bibitem[Johansen et al.(2012)]{johansen} Johansen, A., Davies, M.~B., Church, R.~P., \& Holmelin, V.\ 2012, \apj, 758, 39 

\bibitem[Kipping(2014)]{kipping1} Kipping, D.~M.\ 2014, \mnras, 444, 2263 

\bibitem[Knutson et al.(2014)]{friends_rv} Knutson, H.~A., Fulton, B.~J., Montet, B.~T., et al.\ 2014, \apj, 785, 126 

\bibitem[Lai \& Pu(2016)]{lai} Lai, D., \& Pu, B.\ 2016, arXiv:1606.08855 %
%\bibitem[Laughlin et al.(2011)]{radanom} Laughlin, G., 
%Crismani, M., \& Adams, F.~C.\ 2011, \apjl, 729, L7 

\bibitem[Li \& Winn(2016)]{gongjie_tides} Li, G., \& Winn, J.~N.\ 2016, \apj, 818, 5 

\bibitem[Lissauer et al.(2011a)]{lissauer} 
Lissauer, J. J., Ragozzine, D., Fabrycky, D. C. et al. 2011, ApJS, 197, 8 

\bibitem[Lissauer et al.(2011b)]{systems2010b} Lissauer, J.~J., Fabrycky, D.~C., Ford, E.~B., et al.\ 2011, \nat, 470, 53 

\bibitem[Marcy et al.(2014)]{systems2014c} Marcy, G.~W., Isaacson, H., Howard, A.~W., et al.\ 2014, \apjs, 210, 20 

\bibitem[Mazeh et al.(2015)]{mazeh} Mazeh, T., Perets, H.~B., McQuillan, A., \& Goldstein, E.~S.\ 2015, \apj, 801, 3 

\bibitem[Mills \& Fabrycky(2016)]{misalign108} Mills, S.~M., \& Fabrycky, D.~C.\ 2016, arXiv:1606.04485 

\bibitem[Moriarty \& Ballard(2015)]{johnm} Moriarty, J., \& Ballard, S.\ 2015, arXiv:1512.03445 


\bibitem[Morton \& Winn(2014)]{tim} Morton, T.~D., \& Winn, J.~N.\ 2014, \apj, 796, 47 

\bibitem[Mustill et al.(2016)]{priors_mustill} Mustill, A.~J, Davies, M.~B, \& Johansen, A.\ 2016, arXiv:1609.08058 


\bibitem[Murray \& Dermott(1999)]{md} Murray, C.~D., \& Dermott, S.~F.\ 1999, Solar system dynamics by Murray, C.~D., 1999,  

%\bibitem[Muirhead et al.(2012)]{kepchar1} Muirhead, P.~S., 
%Hamren, K., Schlawin, E., et al.\ 2012, \apjl, 750, L37 
\bibitem[Neveu-VanMalle et al.(2016)]{marion} Neveu-VanMalle, M., Queloz, D., Anderson, D.~R., et al.\ 2016, \aap, 586, A93 

\bibitem[Ngo et al.(2015)]{friends1} Ngo, H., Knutson, H.~A., Hinkley, S., et al.\ 2015, \apj, 800, 138 


\bibitem[Petigura et al.(2013)]{petigura} Petigura, E.~A., Howard, A.~W., \& Marcy, G.~W.\ 2013, Proceedings of the National Academy of Science, 110, 19273 

\bibitem[Petrovich et al.(2014)]{petrovich} Petrovich, C., Tremaine, S., \& Rafikov, R.\ 2014, \apj, 786, 101 

\bibitem[Pordes(2008)]{osg1} Pordes, R. et al. 2007, J. Phys. Conf. Ser. 78, 012057

\bibitem[Pu \& Wu(2015)]{spacing} Pu, B., \& Wu, Y.\ 2015, \apj, 807, 44 


\bibitem[Quintana et al.(2007)]{quintana}
Quintana, E. V., Adams, F. C., Lissauer, J. J., \& Chambders, J. E. 2007, ApJ, 660, 807 

\bibitem[Quintana et al.(2014)]{systems2014} Quintana, E.~V., Barclay, T., Raymond, S.~N., et al.\ 2014, Science, 344, 277 

\bibitem[Rowe et al.(2014)]{systems2014b} Rowe, J.~F., Bryson, S.~T., Marcy, G.~W., et al.\ 2014, \apj, 784, 45 

\bibitem[Rogers(2015)]{notrocky} Rogers, L.~A.\ 2015, \apj, 801, 41 

\bibitem[Sanchis-Ojeda et al.(2015)]{roberto} Sanchis-Ojeda, R., Winn, J.~N., Dai, F., et al.\ 2015, \apjl, 812, L11 


\bibitem[Sfiligoi(2009)]{osg2} Sfiligoi, I., Bradley, D. C., Holzman, B., Mhashilkar, P., Padhi, S. and Wurthwein, F. (2009). 2009 WRI World Congress on Computer Science and Information Engineering, Vol. 2, pp. 428-432. 

\bibitem[Spalding et al.(2014)]{chris1} Spalding, C., Batygin, K., \& Adams, F.~C.\ 2014, \apjl, 797, L29 

\bibitem[Spalding et al.(2016)]{secular3} Spalding, C., Batygin, K., \& Adams, F.~C.\ 2016, \apj, 817, 18 


%\bibitem[Swift et al.(2013)]{kepler32} Swift, J.~J., Johnson, J.~A., Morton, T.~D., et al.\ 2013, \apj, 764, 105 

\bibitem[Towns et al.(2014)]{xsede1} Towns, J., Cockerill, T., Dahan, M., et al. 2014, Computing in Science \& Engineering, 16, 5, pp. 62-74

\bibitem[Uehara et al.(2016)]{uehara} Uehara, S., Kawahara, H., Masuda, K., Yamada, S., \& Aizawa, M.\ 2016, \apj, 822, 2 

\bibitem[Van Laerhoven \& Greenberg(2012)]{secular2} Van Laerhoven, C., \& Greenberg, R.\ 2012, Celestial Mechanics and Dynamical Astronomy, 113, 215 

\bibitem[Vanderburg et al.(2016)]{fiveplanet} Vanderburg, A., Becker, J.~C., Kristiansen, M.~H., et al.\ 2016, \apjl, 827, L10 


\bibitem[Volk \& Gladman(2015)]{volk} Volk, K., \& Gladman, B.\ 2015, \apjl, 806, L26 

\bibitem[Wang et al.(2015)]{legacy} Wang, J., Fischer, D.~A., Barclay, T., et al.\ 2015, \apj, 815, 127 


\bibitem[Weiss 
\& Marcy(2014)]{weiss} Weiss, L.~M., \& Marcy, G.~W.\ 2014, \apjl, 783, L6 

\bibitem[Winn et al.(2010)]{temp6250} Winn, J.~N., Fabrycky, D., Albrecht, S., \& Johnson, J.~A.\ 2010, \apjl, 718, L145 


\bibitem[Wolfgang et al.(2015)]{angie} Wolfgang, A., Rogers, 
L.~A., \& Ford, E.~B.\ 2015, arXiv:1504.07557 

\bibitem[Ziegler et al.(2016)]{ao16} Ziegler, C., Law, N.~M., Baranec, C., et al.\ 2016, \procspie, 9909, 99095U 


\end{thebibliography}
\end{document}